\documentclass[pre,reprint,superscriptaddress,twocolumn,floatfix,aps]{revtex4-2}

\usepackage{microtype}
\usepackage[utf8]{inputenx}
\usepackage{physics}                            % Per i simboli di fisica
\usepackage{bm}									% Per i simboli in grassetto
\usepackage{enumerate}							% Per personalizzare gli elenchi numerati
\usepackage{amsthm}								% Per teoremi e definizioni
\usepackage{comment}							% Per i commenti su più righe
\usepackage[caption=false]{subfig}
\usepackage{amssymb}							% Per i simboli degli insiemi matematici
\usepackage{t1enc}								% Per le virgolette caporali
\usepackage{mathtools}							% Per i sistemi di equazioni
\usepackage{tensor}								% Per gli indici alti e bassi
\usepackage{wrapfig}							% Per le figure allineate a lato
\usepackage{listings}							% Per le linee di codice
\usepackage[usenames,dvipsnames]{xcolor}								% Per parole o blocchi di testo colorati
\usepackage{hyperref}
\hypersetup{colorlinks=true}
\usepackage[capitalize]{cleveref}               % Clever referencing
\usepackage{graphicx,color,xcolor}
\usepackage{amsfonts,amsmath}
\usepackage{xstring}                            % For clever citations
\usepackage{tikz}

\newcommand{\cor}[1]{\mathcal{#1}}									% Calligrafico
\newcommand{\T}[1]{\text{#1}}										% Testo nelle equazioni
\newcommand{\dslash}[1]{\frac{\dd[d]{#1}}{(2\pi)^d}}                % Momentum integral
\def \v {^\T{(v)}}
\def \q {_{\vb q}}
\newcommand{\n}{\nonumber}
\newcommand{\ccite}[1]{\IfSubStr{#1}{,}{Refs.~}{Ref.~}\cite{#1}}

\newcommand{\K}{\hat{\mathcal K}}

\newcommand{\rev}[1]{#1}

\begin{document}
\title{Structure and dynamics of a Rouse polymer in a fluctuating correlated medium}
\author{Pietro Luigi Muzzeddu}
\thanks{Corresponding author}
\email{pmuzzedd@sissa.it}
\affiliation{Department of Biochemistry, University of Geneva, 1205 Geneva, Switzerland}
\affiliation{SISSA --- International School for Advanced Studies and INFN, via Bonomea 265, 34136 Trieste, Italy}
\author{Davide Venturelli}
\affiliation{Laboratoire de Physique Th\'eorique de la Mati\`ere Condens\'ee, CNRS/Sorbonne Universit\'e, 4 Place Jussieu, 75005 Paris, France}
\affiliation{Laboratoire Jean Perrin, CNRS/Sorbonne Universit\'e, 4 Place Jussieu, 75005 Paris, France}
\author{Andrea Gambassi}
\affiliation{SISSA --- International School for Advanced Studies and INFN, via Bonomea 265, 34136 Trieste, Italy}

\begin{abstract}
We study the static and dynamical properties of a harmonically confined Rouse polymer coupled to a fluctuating correlated medium, which affect each other reciprocally during their stochastic evolution. 
The medium is modeled by a scalar Gaussian field 
%$\phi(\mathbf{x},t)$, 
which can feature modes with slow relaxation and long-range spatial correlations.
We show that these modes affect the long-time behavior of the %center-of-mass 
average position of the center of mass of the polymer, which, after a displacement, turns out to relax algebraically towards its equilibrium value. %position. 
This is a manifestation of the non-Markovian nature of the effective evolution of the position of the center of mass, once the degrees of freedom of the medium have been integrated out.
%\dav{We attribute this decay to the behavior of the memory kernel in the non-Markovian effective evolution equation of the center of mass, after integrating out the medium’s degrees of freedom.}
%\dav{We rationalize this decay in terms of that of the memory kernel that features in the non-Markovian, effective evolution equation of the center of mass, once the degrees of freedom of the medium have been integrated out.}
%\plm{We prove that the exponents of such relaxation can be related to the non-Markovian nature of the effective evolution of the center of mass and, more precisely, to the power-law decay of the memory kernel.}
In contrast, we show that the coupling to the medium speeds up the relaxation of higher Rouse modes. 
We further characterize the typical size of the polymer as a function of its polymerization degree and of the correlation length of the medium, particularly when the system is driven out of equilibrium via the application of a constant external driving force. 
Finally, we study the response of a linear polymer to a tensile force acting on its terminal monomers.
\end{abstract}

\maketitle

\tableofcontents

\section{Introduction}
\label{sec:intro}

Understanding the behavior of polymeric macromolecules dispersed in a fluid environment is of crucial importance for advances in biomedical applications~\cite{kirillova2019shape, hoffman2013stimuli, jeong2002lessons}, the design and development of smart materials~\cite{mukherji2020smart, aguilar2019smart, galaev1999smart, kumar2007smart}, and the comprehension of several biological processes~\cite{brangwynne2015polymer, mackintosh1995elasticity}. In most situations, polymeric molecules are in contact with complex heterogeneous and correlated media, and their behavior is significantly affected by 
%this 
the mutual interaction. For example, this is the case of polymers embedded in composite fluids~\cite{tanaka2008temperature}, porous media~\cite{viovy2000electrophoresis}, biological tissues and cellular interiors~\cite{chiariello2016polymer, wen2011polymer}. 
Over the last decades, particular emphasis has been put 
%in 
on investigating structural properties of polymeric chains in binary liquid mixtures displaying 
%statistical 
spatio-temporal correlations~\cite{DeGennes1976,deGennes_1980,Magda1988,Mukherji2014,zheng2018unusual, venkatesu2006polymer, araki2016conformational, to1998abnormal}. 
The typical length scale of such correlations depends on the distance 
%of the binary mixture 
from the critical point of demixing (occurring at a temperature $T_c$)
of the binary mixture, 
and it potentially diverges when the mixture is poised at the critical point. 

The first theoretical analysis of this problem was conducted by De Gennes and Brochard~\cite{DeGennes1976, deGennes_1980}, who showed that a polymer dispersed in a binary liquid mixture would first collapse into a globule-like configuration as the solvent approaches the demixing transition, to eventually re-expand at the critical point itself. 
The polymer collapse is explained by assuming that the better solvent of the binary mixture forms a %wetting
layer around the polymer, screening the excluded volume repulsion and thus resulting in an effective attraction. 
The range of these induced interactions caused by the fluctuating medium is given by the correlation length of the latter~\cite{DeGennes1976, vilgis1993conformation}. In particular, the effective interactions experienced by the polymer become scale-free 
when the underlying medium is critical. 
When the correlation length of the binary mixture exceeds the typical size of the polymer, the latter is effectively immersed in a droplet enriched with the better solvent, and thus it swells again to its size in a pure solvent. Later, additional studies were performed, relying on self-consistent perturbative schemes~\cite{dua1999polymer, edwards1979size} and field-theoretic methods~\cite{vilgis1993conformation, stapper1998behavior, negadi1999mean, negadi2000dynamic}. 
The analytical predictions of the conformational properties of a polymer in a near-critical solvent have been verified by on-lattice Monte Carlo simulations~\cite{Magda1988, vasilevskaya1998conformation}, and multiscale simulation methods based on density functional theory~\cite{sumi2009critical}. From an experimental perspective, the effect of a correlated environment on the structure of a tracer chain has been explored using dynamic light scattering~\cite{theobald1997evidence, to1998polymer} and small-angle neutron scattering~\cite{he2012partial}.
%
%\ag{Is there anything relevant in these studeis for pur work?}

In this work, we study conformational and dynamical properties of a polymer chain interacting with a correlated medium, which is described by a thermally fluctuating scalar order parameter field $\phi(\bm{x},t)$. The latter evolves according to a purely dissipative or a conserved relaxational dynamics, 
and interacts with the polymer via a translationally invariant linear coupling~\cite{venturelli2022nonequilibrium, venturelli2022inducing, venturelli2023memory, venturelli2024heat, venturelli2022tracer, demery2011perturbative, demery2013diffusion, Dean2011}. Due to this interaction, the polymer and the field affect each other dynamically, in such a way that detailed balance is satisfied along their stochastic evolution. To make contact with the case of a polymer chain dispersed in a binary liquid solvent mixture, the scalar field $\phi(\bm{x},t)$ can be interpreted as the relative concentration of the two species in the mixture
(although in this analogy we neglected for simplicity hydrodynamic effects, and other slow variables that should be taken into account when describing real fluids~\cite{hohenberg1977theory}).
However, our analysis below is more general, and does not rely on any specific interpretation of the order parameter $\phi(\bm{x},t)$. 
%With the help of two theoretical methods 
Using two distinct approaches based on the linear-response theory 
%and 
or the weak-coupling approximation, we derive theoretical predictions on conformational and relaxation properties of the chain. While, as anticipated above, many static properties of a polymer in a correlated environment have been widely studied in the past, the effect of the field-mediated forces on its non-equilibrium dynamics, which we address in this work, is much less explored.

To this end, after describing the model in Section~\ref{sec:model} and analyzing the field-induced interactions in Section~\ref{subsec:stat_dist}, we derive in Section~\ref{sec:relaxation} a linearized effective equation of motion for the polymer,
%and study the dynamical relaxation of its internal structure and of its center of mass towards the rest position in a confining potential.
finding that the evolution of the coordinates of its center of mass is governed by a generalized Langevin equation (GLE), with a memory kernel that depends on the correlation length of the medium. In particular, when the field 
%order parameter 
supports modes with slow relaxation, we show that the memory kernel decays algebraically with time, and thus memory effects are more prominent.
Then, we study the dynamical relaxation of the internal structure of the polymer, and of its center of mass towards the rest position in a confining potential.
In particular, we investigate the extent to which the correlated medium affects this relaxation process, especially close to the critical point. Next, in 
Section~\ref{sec:polymer_size} 
%\cref{sec:typical_no_f} 
we study the typical size of the polymer as a function of its polymerization degree and of the correlation length of the medium.
%field. 
In \cref{sec:force-extension} we analyze the response of a linear polymer to a tensile force 
applied to its terminal monomers.
Finally, in \cref{sec:moving-trap} we consider the case in which the polymer is subject to a constant external driving, and predict how the nonequilibrium field-mediated forces affect the typical size of the polymer in that case. 
All analytical predictions are tested %with 
using numerical simulations.

%%
%%
%%
%%%%%%%%%%%%%%%%%%%%%%%%%%%%%%%%%%%%%%%%%%%%%%%%%%%%%%%%%
%%%%%%%%%%%%%%%%%%%%%%%%%%%%%%%%%%%%%%%%%%%%%%%%%%%%%%%%%
\begin{figure}[t]
    \includegraphics[width=0.9\linewidth]{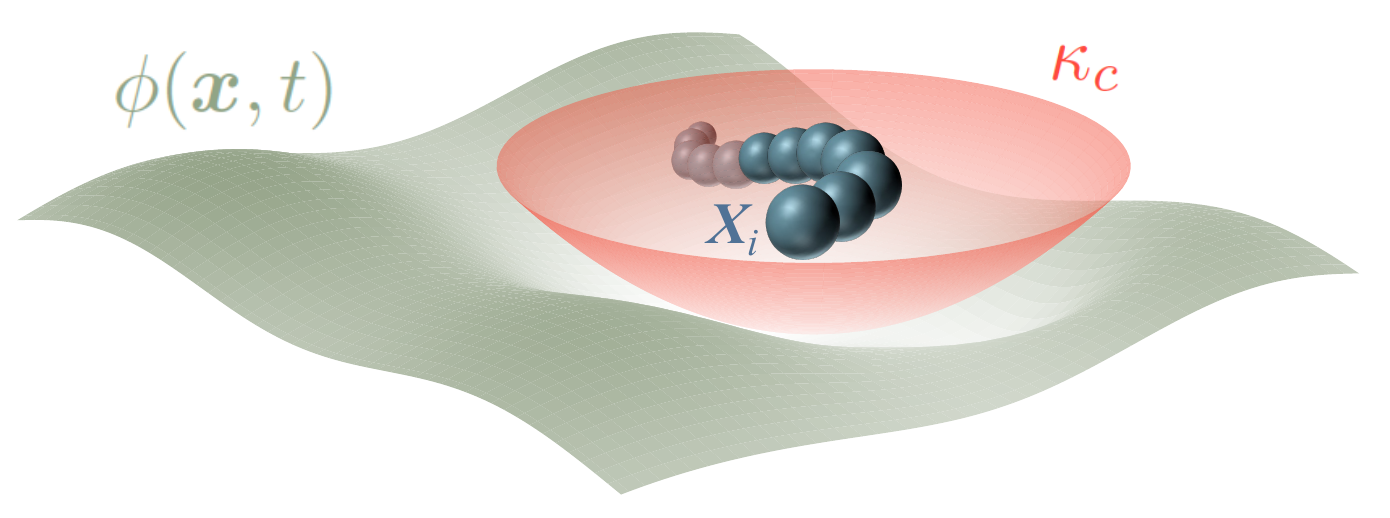}
    \caption{Schematic illustration of the model, consisting of a (linear) polymer chain coupled to a thermally fluctuating order parameter field $\phi(\bm{x},t)$, and confined by a harmonic potential with stiffness $\kappa_c$. }
    \label{fig:sketch}
\end{figure}
%%%%%%%%%%%%%%%%%%%%%%%%%%%%%%%%%%%%%%%%%%%%%%%%%%%%%%%%%

\section{The model}
\label{sec:model}
The system 
investigated in this work, schematically illustrated in Fig.~\ref{fig:sketch},
consists of an ideal harmonic (Rouse) chain in $d$ spatial dimensions, composed by $N$ monomers with positions $\{\bm{X}_i(t)\}_{i=0}^{N-1}$, and a fluctuating order parameter field $\phi(\bm{x},t)$. The structure of the internal interactions between the monomers is encoded in the connectivity matrix $M_{ij}$, with $i,j\in\{0,1,\ldots, N-1\}$, so that the Hamiltonian of the Rouse polymer is given by:
\begin{equation}
    \mathcal{H}_0=\frac{\kappa}{2} \sum_{i,j=0}^{N-1} M_{ij}\bm{X}_i \cdot \bm{X}_j + \frac{\kappa_c}{2}\sum_{i=0}^{N-1} \bm{X}_i^2,
    \label{eq:hamiltonian_chain}
\end{equation}
where $\kappa$ denotes the stiffness of the pairwise attractions between the sub-units of the chain, while $\kappa_c$ is the elastic constant of the external harmonic confinement, in case this is present. 
Note that, in this work, we neglect energetic contributions coming from the bending of the polymeric chain, as well as excluded-volume interactions leading to steric hindrance effects, as we aim to investigate the extent to which the spatio-temporal correlations in the underlying medium affect the structural \rev{and dynamical} properties of the simplest possible polymer model. For the fluctuating scalar field $\phi(\bm{x},t)$ we take a Gaussian Hamiltonian
\begin{equation}
    \mathcal{H}_{\phi}=\int \dd^d \bm{x} \left[ \frac{1}{2}(\nabla \phi)^2 + \frac{r}{2}\phi^2 \right],
    \label{eq:hamiltonian_field}
\end{equation}
where the parameter $r\geq0$ controls the deviation of the field from criticality, and determines its correlation length $\xi_\phi=r^{-1/2}$. Analogously to, e.g., Refs.~\cite{venturelli2022nonequilibrium, basu2022dynamics, venturelli2022inducing, venturelli2023memory,demery2011perturbative}, 
the coupling between the polymer and the field is chosen to be linear and translationally invariant, and it is given by:
\begin{equation}
    \mathcal{H}_{\text{int}}=- \lambda \sum_{i=0}^{N-1} \sigma_i \int \dd^d\bm{x}\,  \phi(\bm{x}) V(\bm{x}-\bm{X}_i),
    \label{eq:interaction_hamiltonian}
\end{equation}
where $\lambda>0$ denotes the coupling strength, $V(\bm{x})>0$ is the interaction potential, and $\{ \sigma_i\}_{i=0}^{N-1}$ is a set of $N$ 
%spin-like 
binary variables 
%with 
$\sigma_i \in \{-1,+1 \}$. 
This means that the energetically favored configurations are those where the field assumes larger (smaller) values in the spatial proximity of the monomers with positive (negative) interaction coupling $\lambda \sigma_i$. For simplicity, we focus on isotropic interaction potentials characterized by a single length scale $R$, such as
\begin{equation}
    V(\bm{x})=(2 \pi R^2)^{-d/2} 
    %\exp\left\{
    \exp(-    \frac{\bm{x}^2}{2 R^2})
    %\right\}
    .
    \label{eq:Gaussian_interaction_potential}
\end{equation}
Here $R$ represents the characteristic length scale of interaction between the field $\phi(\bm{x},t)$ and each monomer in the chain, and it might be interpreted as the typical size of a monomer, assuming that they are all equal. In the following, we will denote the total Hamiltonian as
\begin{equation}
    \mathcal{H}=\mathcal{H}_0 + \mathcal{H}_{\phi} + \mathcal{H}_{\text{int}}.
    \label{eq:total_hamiltonian}
\end{equation}
As typically done for biomolecules in solution, we assume that viscous forces dominate over inertial effects, and we model the equation of motion of the polymer with the following set of overdamped Langevin equations:
\begin{align}
    \dot {\bm{X}}_i(t)&=-\nu \nabla_{\bm{X}_i} \mathcal{H} + \bm \xi_i(t)
    \label{eq:monomers_dynamics}
    \\&=-\nu \kappa \sum_j M_{ij}\bm{X}_j - \nu \kappa_c \bm{X}_i + \nu \lambda \sigma_i \bm{{\rm f}}(\bm{X}_i) + \bm{\xi}_i(t), \n
\end{align}
where $\nu$ is the monomer mobility, and the force $\bm{{\rm f}}(\bm{X}_i)$ exerted by the field on the $i$-th monomer reads
\begin{equation}
    \bm{{\rm f}}(\bm{X}_i,\phi,t) \equiv  -\int \dd^d\bm{x}\, \phi(\bm{x})\nabla_{\bm{x}}V(\bm{x}-\bm{X}_i).
    \label{eq:field_induced_force}
\end{equation}
Moreover, the polymer is in contact with a thermal bath at temperature $T$, the effect of which is accounted for by the set of zero-mean independent Gaussian white noises 
%$\left\{\bm{\xi}_i(t)\right\}$, 
$\bm{\xi}_i(t)$, 
with correlations
\begin{equation}
    \langle \xi_{i}^\alpha(t) \xi_{j}^\beta(s)\rangle=2\nu T\delta_{ij}\delta_{\alpha \beta}\delta(t-s).
    \label{eq:corr_xi}
\end{equation}
Here, $\alpha,\beta \in \{0,1,...,d-1\}$ denote the spatial directions associated to the canonical basis $\{\hat{\bm{e}}_\alpha\}$, with $\hat{\bm{e}}_0$ corresponding to the $x$-axis.
Being the system at thermal equilibrium, the amplitude of the noise is chosen such that fluctuations and dissipation are related by the Einstein relation. The stochastic dynamics in Eq.~\eqref{eq:monomers_dynamics} can be rewritten within the Rouse domain~\cite{doi1988theory} by introducing the Rouse modes 
\begin{equation}
    \bm{\chi}_i=\sum_{j=0}^{N-1} \varphi_{ij} \bm{X}_j,
    \label{eq:Rouse_transformation}
\end{equation}
where the orthogonal transformation $\bm{\varphi}$ diagonalizes the connectivity matrix $\bm{M}$, and is chosen such that its rows are normalized to unity.
We recall that the connectivity matrix $M_{ij}$ is symmetric, and such that the sum over its rows (or columns) vanishes by construction~\cite{rubinstein2003polymer}.
The transformation defined above implies that the %$0$-order 
Rouse mode $\bm{\chi}_0$ is related to the  center of mass $\bm{X}_{\text{com}}$ of the polymer by the simple relation $\bm{\chi}_0=\sqrt{N} \bm{X}_{\text{com}}$, while the higher-order Rouse modes contain information about the internal structure of the chain. 
When the coupling to the field is switched off, i.e., for $\lambda=0$, the Rouse modes $\bm{\chi}_i$ are decoupled and their time evolution is governed by independent Ornstein-Uhlenbeck processes with inverse relaxation times
\begin{equation}
    \tau_{i}^{-1} \equiv \gamma_i + \gamma_c \equiv \tilde{\gamma}_i.
    \label{eq:relaxation_rates_polymer}
\end{equation}
Here $\gamma_c=\nu \kappa_c$ is the inverse characteristic time scale introduced by the harmonic confinement, while $\gamma_i=\nu \kappa m_i$ 
denotes the inverse relaxation time of the $i$-th Rouse mode $\bm{\chi}_i$ of an unconfined chain, which depends on the corresponding eigenvalue $m_i$ of the connectivity matrix $\bm{M}$. However, when the polymer interacts with $\phi(\bm{x},t)$, the field-induced forces couple the Rouse modes yielding the following dynamics:
\begin{equation}
         \dot{\bm{\chi}}_i=-\tilde{\gamma}_i \bm{\chi}_i + \nu \lambda \sum_{j=0}^{N-1} \varphi_{ij} \sigma_j \bm{{\rm f} }(\bm{X}_j,\phi) +  \bm{\eta}_i(t),
    \label{eq:dynamics_rouse_modes}
\end{equation}
where the noise $\bm{\eta}_i$ has the same statistics as $\bm{\xi}_i$, and the monomer position $\bm{X}_j$ can be rewritten as a linear combination of Rouse modes by means of the inverse transformation $\bm{\varphi}^{-1}$.

The field $\phi(\bm{x},t)$ is further assumed to evolve according to the relaxational dynamics~\cite{tauber2014critical}
\begin{align}
&\partial_t \phi(\bm{x},t)=-D(i\nabla)^a \frac{\delta \mathcal{H}}{\delta \phi(\bm{x},t)} + \zeta(\bm{x},t) \label{eq:field_dynamics}\\
&=-D(i\nabla)^a \left[(r-\nabla^2)\phi - \lambda \sum_{i=0}^{N-1} \sigma_i V(\bm{x}-\bm{X}_i) \right]+ \zeta(\bm{x},t), \n
\end{align}
where $D$ denotes the mobility of the field. 
The parameter $a$ takes the value $a=2$ if the field is locally conserved along its dynamics, whereas $a=0$ if the field order parameter does not satisfy any conservation law. 
The two cases $a=0$ and $a=2$ correspond to the so-called model~A and model~B dynamics within the Gaussian approximation, respectively~\cite{hohenberg1977theory, tauber2014critical}. 
For $a=2$, Eq.~\eqref{eq:field_dynamics} takes the form of a continuity equation $\partial_t \phi(\bm{x},t)=-\nabla \cdot \bm{\mathcal{\bm{J}}}(\bm{x},t)$, with $\bm{\mathcal{\bm{J}}}(\bm{x},t)$ a fluctuating flux, as expected due to the underlying conservation law. The zero-mean Gaussian white noise field $\zeta(\bm{x},t)$ in Eq.~\eqref{eq:field_dynamics} is characterized by the correlations
\begin{equation}
    \langle \zeta(\bm{x},t) \zeta(\bm{y},s) \rangle=2DT(i\nabla)^a \delta^d(\bm{x}-\bm{y})\delta(t-s).
    \label{eq:correlations_noise_field}
\end{equation}
The amplitude of the noise field is proportional to the field mobility $D$ and to the temperature $T$, as the polymer and the field are assumed to be in contact with the same thermal bath in equilibrium. The stochastic dynamics of the field can be conveniently written in Fourier space as
\begin{equation}
    \dot \phi_{\bm{q}}=-\alpha_{\bm{q}}\phi_{\bm{q}} + D\lambda V_{\bm{q}}q^a \sum_{j=0}^{N-1} \sigma_j \mathrm{e}^{-i\bm{X}_j\cdot \bm{q}} + \zeta_{\bm{q}}(t),
    \label{eq:field_dynamics_fourier}
\end{equation}
where we introduced 
\begin{equation}
    \alpha_{\bm{q}} \equiv D q^a (r + q^2),
\label{eq:def-alphaq}
\end{equation}
and $\zeta_{\bm{q}}(t)$ satisfies
\begin{equation}
    \langle \zeta_{\bm{q}}(t)\zeta_{\bm{q}'}(t') \rangle= 2DT (2\pi)^d q^a\delta^d(\bm{q}+\bm{q}')\delta(t-s).
    \label{eq:correlations_noise_field_fourier}
\end{equation}
Similarly to Ref.~\cite{venturelli2022nonequilibrium}, in the case of model A dynamics the field mode $\phi_{\bm{q}=\bm{0}}$ may take arbitrarily large values as criticality is approached. However, this effect is irrelevant for what concerns the equation of motion of the polymer, as it can be easily seen  by rewriting the field-induced force $\bm{{\rm f}}$ in Eq.~\eqref{eq:field_induced_force} as 
\begin{equation}
     \bm{{\rm f}}(\bm{X}_j,\phi,t) =\int \dslash{q} \,i\bm{q}\phi_{\bm{q}}V_{-\bm{q}}\mathrm{e}^{{\rm i}\bm{q}\cdot \bm{X}_j},
    \label{eq:field_mediated_forces}
\end{equation}
where the mode with $\bm{q}=\bm{0}$ does not contribute. 
In a more realistic model, one would need to counteract this growth 
by adding a suitable chemical potential --- e.g., $\cor H_\phi \mapsto \cor H_\phi + \lambda N \int \dd{\bm{x}} \phi(\bm{x}) $. 
In the absence of the interaction with the particle, i.e., for $\lambda=0$, the field modes $\phi_{\bm{q}}$ evolve according to independent Ornstein-Uhlenbeck processes with relaxation times $\tau_{\phi}(\bm{q})=\alpha_{\bm{q}}^{-1}$. In particular, this implies that the $\bm{q}=\bm{0}$ mode features a diverging relaxation time in the case of critical model A dynamics, which is responsible for the so-called critical slowing down~\cite{tauber2014critical}. In the case of model~B dynamics, this slow relaxation occurs even away from criticality as a consequence of the conservation law, whereby fluctuations at large length scales relax on arbitrarily long time scales.  

%%%%%%%%%%%%%%%%%%%%%%%%%%%%%%%%%%%%%%%%%%%%%%%%%%%%%%%%%
%%%%%%%%%%%%%%%%%%%%%%%%%%%%%%%%%%%%%%%%%%%%%%%%%%%%%%%%%
\begin{figure*}[t]
    %\centering
    %\begin{subfigure}{0.49\linewidth}
    \subfloat{
        \includegraphics
        %[width=\linewidth, height=6.2cm]
        [width=0.49\linewidth, height=6.2cm]
        {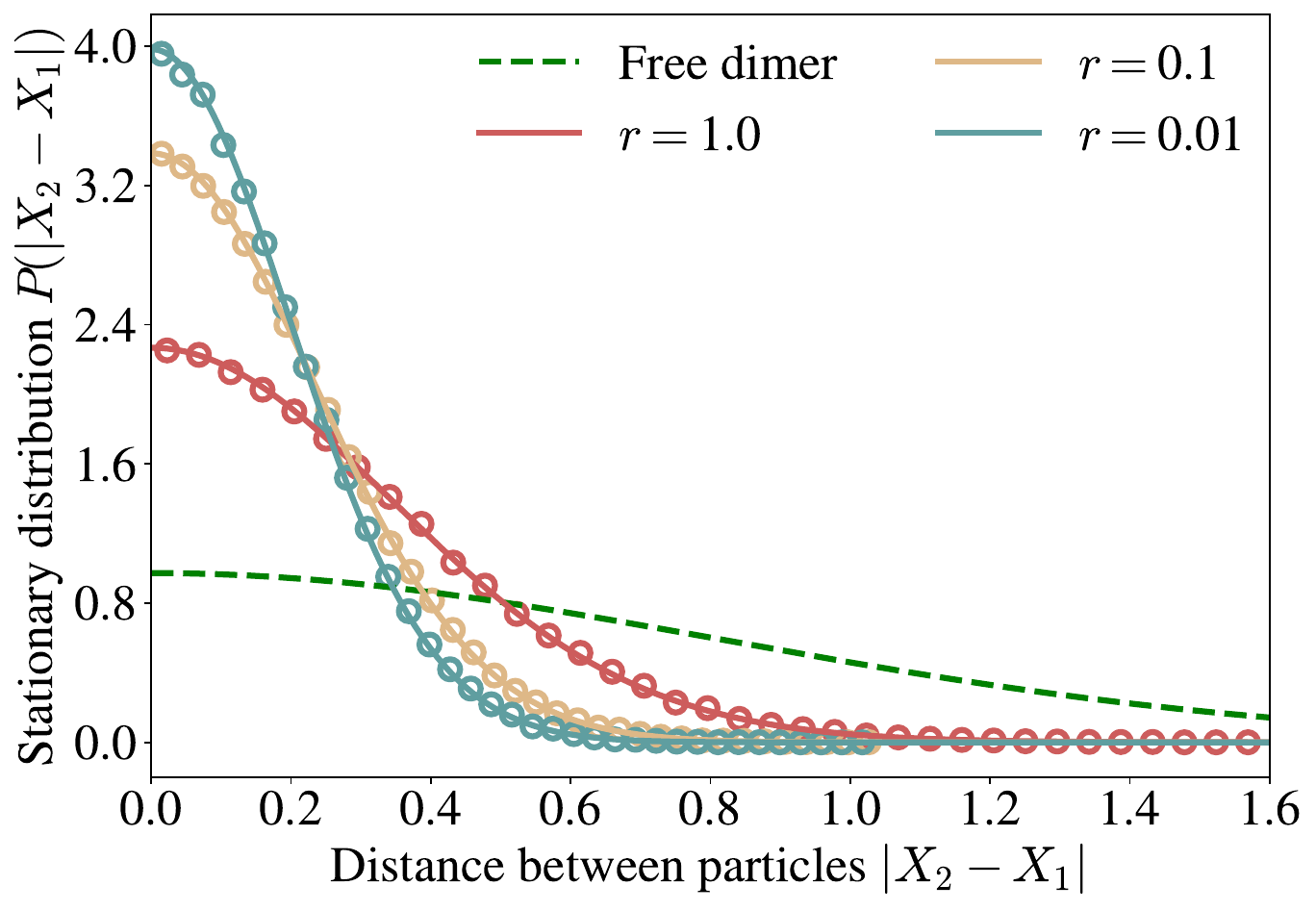}
        %\put(-215,101){(a)}
        \label{aa}
    }
    %\end{subfigure}
    %\hfill
    %\begin{subfigure}{0.49\linewidth}
    \subfloat{
        \includegraphics
        %[width=\linewidth, height=6.2cm]
        [width=0.49\linewidth, height=6.2cm]
        {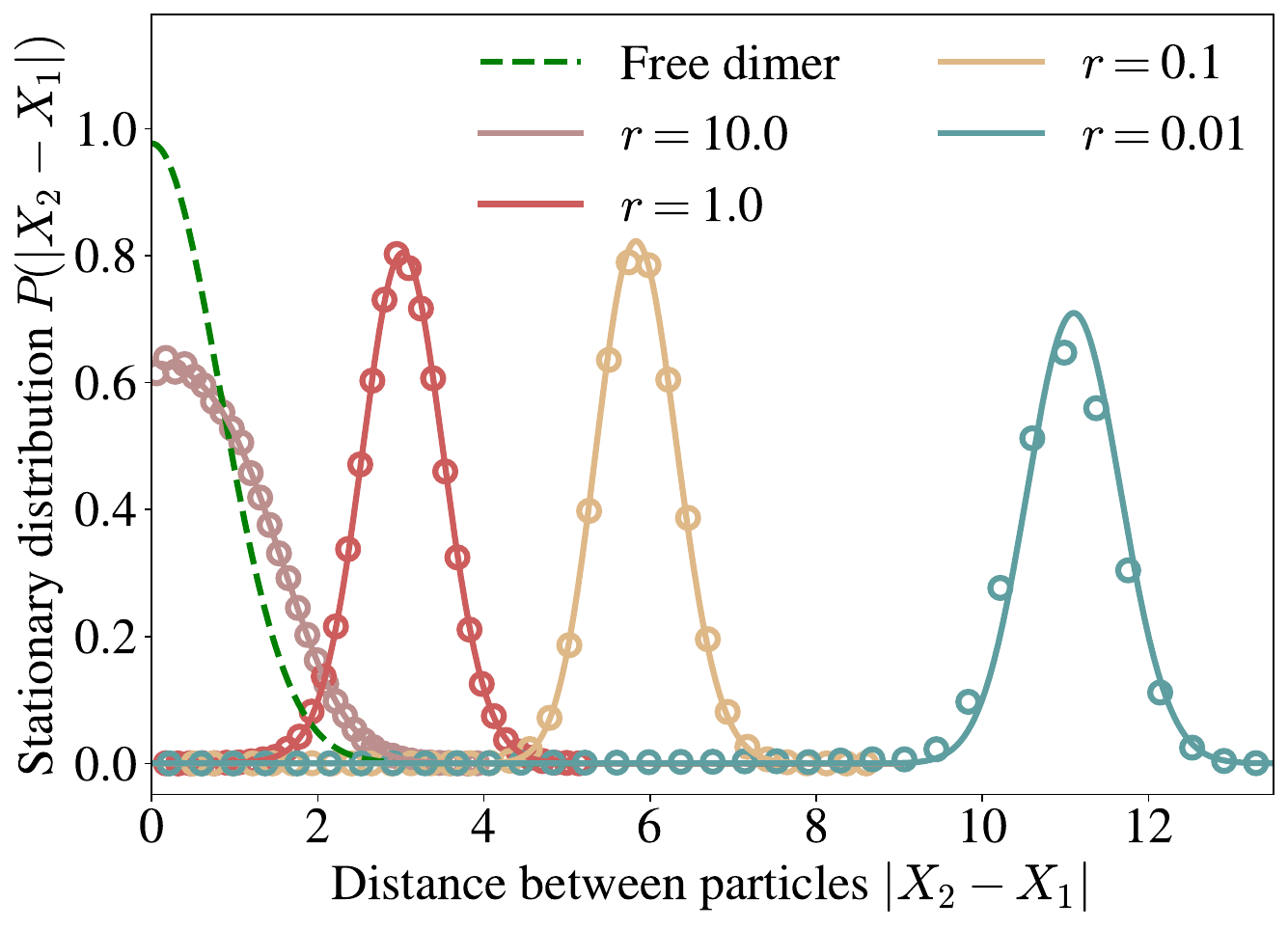}
        \label{ab}
    }
    %\end{subfloat}
    \caption{Equilibrium distribution of the distance $|X_2-X_1|$ between the two sub-units of a dimeric molecule at positions $X_{1,2}$, in $d=1$ and in the absence of confinement. Solid lines represent the theoretical prediction in Eq.~\eqref{eq:marginal_distribution_polymer} specialized to the case $N=2$, whereas symbols are obtained with molecular dynamics simulations (see App.~\ref{sec:simulations} for details). Left panel: the monomers have the same interaction coupling with the fluctuating field, i.e., $\sigma_1=\sigma_2$. This induces a collapse of the dimer, which is increasingly more effective as the field approaches criticality (i.e., as $r\to 0$). Right panel: the monomers have opposite interaction couplings, i.e., $\sigma_1= - \sigma_2$, resulting in a repulsive interaction and therefore in a stretching of the dimer.   
    %Simulation parameters: $\kappa=1$, $\nu=1$, $T=1$, $D=1$, $R=1$, $\lambda=10$. 
    In the simulation we chose $\lambda=10$, while all other parameters were set to unity.
    }
    \label{fig:distance_dimer}
\end{figure*}
%%%%%%%%%%%%%%%%%%%%%%%%%%%%%%%%%%%%%%%%%%%%%%%%%%%%%%%%%

\section{Induced dynamical interactions on the polymer}
\label{sec:interaction_polymer}

To investigate the effect of the correlated medium on the structural and dynamical properties of the polymer, we analyze the behavior of its center of mass $\bm{X}_{\text{com}}=\bm{\chi}_0/\sqrt{N}$ and of its internal structure described by the Rouse modes $\bm{\chi}_j$ with $j\geq 1$. 
By computing their amplitudes $\langle \bm{\chi}_j^2\rangle$, which constitute the so-called power spectrum of the chain, we can also determine the typical size of the polymer given by its mean-square  
%
%\ag{or "mean-square"?}
%
gyration radius. Note that the definition of the Rouse modes in terms of the coordinates of the monomers (see Eq.~\eqref{eq:Rouse_transformation}) depends on the polymer connectivity $\bm{M}$, which is kept generic throughout the upcoming derivation.

\subsection{Stationary distribution}
\label{subsec:stat_dist}

Since the stochastic dynamics in Eqs.~\eqref{eq:dynamics_rouse_modes} and~\eqref{eq:field_dynamics} satisfies the detailed balance condition, the stationary joint probability distribution of the polymer and the field follows the Boltzmann distribution
\begin{equation}
    P_{\text{eq}}[\phi,\{\bm{\chi}_j\}] \propto \exp \left\{ -\beta \mathcal{H}[\phi,\{\bm{\chi}_j\}]\right\},
    \label{eq:Boltzmann_distribution_total}
\end{equation}
with $\beta=1/T$. Hence, the marginal distribution $P_{\text{eq}}(\{\bm{\chi}_j\})$ of the polymer can be obtained by integrating out the field $\phi$, and is given by
\begin{equation}
    P_{\text{eq}}(\{\bm{\chi}_j\}) = \int \!\mathcal{D}\phi\; P_{\text{eq}}[\phi,\{\bm{\chi}_j\}]  \propto \exp[ -\beta (\mathcal{H}_0 + \mathcal{H}_{\text{eff}})],
    \label{eq:marginal_distribution_polymer}
\end{equation}
where $\mathcal{H}_0$ is given in Eq.~\eqref{eq:hamiltonian_chain}.
The effect of the field now appears in the form of the effective interaction Hamiltonian 
%$\mathcal{H}_{\text{eff}}$, the expression of which reads
\begin{align}
    &\mathcal{H}_{\rm eff}=-\frac{\lambda^2}{2} \sum_{i,j}\sigma_i\sigma_j\int \dslash{q}  \abs{V_q}^2 \cor C\q \left[\mathrm{e}^{i\vb q \cdot ( \vb X_i-\vb X_j)}-1\right] \n\\&= -\frac{\lambda^2}{2} \sum_{i,j} \sigma_i\sigma_j\int  \dslash{q}  \abs{V_q}^2 \cor C\q \left[\mathrm{e}^{i\vb q \cdot \sum_k ( \varphi_{ki}-\varphi_{kj})\bm \chi_k}-1\right],
    \label{eq:eff_Hamiltonian_polymer}
\end{align}
where 
\begin{equation}
    \cor C\q=
    %\frac{1}{r + \bm{q}^2}
    (r + \bm{q}^2)^{-1}
    \label{eq:static-prop}
\end{equation} 
denotes the Fourier transform of the equilibrium, equal-time correlation function of the field within the Gaussian model (see, e.g., Ref.~\cite{tauber2014critical}). 
Note that in Eq.~\eqref{eq:eff_Hamiltonian_polymer} we adopted the convention for which the value of the effective interaction 
%Hamiltonian is referred 
energy is measured with respect
to the case of perfectly overlapping monomers, such that $\mathcal{H}(\{\bm{\chi}_j=\bm{0}\})=0$. Importantly, the effective Hamiltonian~\eqref{eq:eff_Hamiltonian_polymer} is pairwise additive, as also observed in Ref.~\cite{fournier2021fieldmediated}. 
This absence of multi-body interactions is actually due to the fact that the coupling between the monomers and the field is linear. 
Accordingly, this property is not merely an equilibrium feature, but it carries over to non-equilibrium dynamics, as it was shown in Ref.~\cite{venturelli2022inducing} using path-integral methods.

Interesting conclusions on the effects of the field-induced forces can be drawn by considering the limiting case of point-like monomers, i.e., $R \to 0$, see Eq.~\eqref{eq:Gaussian_interaction_potential}. 
The interaction potential is thus given by $V(\bm{x})=\delta^d(\bm{x})$, and in $d=3$ the effective Hamiltonian $\mathcal{H}_{\text{eff}}$ takes the form of a Yukawa potential
\begin{equation}
    \mathcal{H}_{\text{eff}}=-\sum_{i\neq j}\frac{\lambda^2 \sigma_i \sigma_j}{4 \pi} \frac{\mathrm{e}^{ -|\bm{X}_i-\bm{X}_j|/\xi_\phi}}{|\bm{X}_i-\bm{X}_j|},
    \label{eq:effective_Yukawa}
\end{equation}
with a characteristic decay length which is given by the correlation length $\xi_\phi$ of the field. This result is consistent with what previously found in Refs.~\cite{DeGennes1976, deGennes_1980, vilgis1993conformation} for polymeric molecules in binary fluids. 
In particular, this implies that the internal structure of the polymer is strongly affected by the field, especially when the latter approaches the critical point and the effective interaction is described by a long-range Hamiltonian.
Importantly, the attractive/repulsive nature of the field-induced interaction between two monomers $i$ and $j$ of the chain depends on the sign of their couplings $\lambda \sigma_{i,j}$ with the fluctuating field $\phi$. More precisely, when the two couplings $\sigma_i$ and $\sigma_j$ have the same sign, the monomers are attracted, whereas if they have opposite signs, the monomers repel each other. This effect is explicitly shown in Fig.~\ref{fig:distance_dimer} for the simplest case of a dimeric molecule ($N=2$) in spatial dimension $d=1$. 
There, we plot the stationary distribution of the relative distance $\Delta X$ between the two monomers for various values of the parameter $r$, i.e., for fluctuating fields $\phi$ with different correlation length. Specifically, we show that as $\phi$ approaches the critical point, the typical distance between the monomers decreases (increases) in the case of attractive (repulsive) field-mediated interactions. 
In particular, in the right panel of Fig.~\ref{fig:distance_dimer} we observe that the center of the stationary distribution departs from $\Delta X=0$ for sufficiently small values of $r$, indicating that the field-induced repulsion 
(whose spatial range is set by the correlation length $\xi_\phi=r^{-1/2}$) is eventually exceeding the strength of the harmonic attraction. 
In both panels, the prediction in Eq.~\eqref{eq:eff_Hamiltonian_polymer} is tested against numerical simulations, the details of which are reported in App.~\ref{sec:simulations}.

\subsection{Relaxation towards equilibrium}
\label{sec:relaxation}

To analyze the influence of the correlated fluctuations of the medium on the dynamical properties of the polymer, in this Section we derive the effective equation of motion of the latter by integrating out the fluctuating order parameter $\phi$, similarly to what was done in
Refs.~\cite{demery2011perturbative,demery2014generalized,basu2022dynamics,venturelli2022inducing,venturelli2023memory,venturelli2024heat}. 
For simplicity, we focus on the case in which all monomers 
%are coupled to the field with the same coupling sign $\sigma$, 
have the same field coupling,
i.e., $\sigma_i=\sigma$ for $i=0,\ldots, N-1$,
%$\sigma_0=\sigma_1=...=\sigma_{N-1}=\sigma$, 
implying that all field-mediated interactions in equilibrium are attractive. 

\subsubsection{Effective dynamics of the polymer}

As a first step, we solve exactly the stochastic dynamics of the fluctuating field in Eq.~\eqref{eq:field_dynamics_fourier}, obtaining
\begin{align}
    \phi_{\bm{q}}(t)&=G_{\bm{q}}(t-t_0)\phi_{\bm{q}}(t_0)+\int_{t_0}^t \dd{s} G_{\bm{q}}(t-s)\zeta_{\bm{q}}(s)     \label{eq:exact_solution_field_polymer} \\&+D\lambda \sigma \int_{t_0}^t \dd{s} G_{\bm{q}}(t-s) V_{\bm{q}} q^a \sum_{j=0}^{N-1}\mathrm{e}^{-i\sum_k\varphi^{-1}_{jk}\bm{q}\cdot \bm{\chi}_k(s)}, \n
\end{align}
where 
\begin{equation}
    G_{\bm{q}}(t)=\Theta(t)\mathrm{e}^{-\alpha_{\bm{q}}t}
    \label{eq:propagator}
\end{equation}
%G_{\bm{q}}(t)=\Theta(t)\mathrm{e}^{-\alpha_{\bm{q}}t}$ 
denotes the dynamic response function of the field~\cite{tauber2014critical},
and $t_0\leq t$ is the initial time, at which the field has an initial configuration $\phi_{\bm{q}}(t_0)$.
We then substitute the solution above
into the evolution equation~\eqref{eq:dynamics_rouse_modes} of the Rouse modes,
yielding the following effective non-Markovian dynamics for the polymer:
\begin{align}
    \dot{\bm{\chi}}_j=&\, \int_{t_0}^t \dd{s} \bm{F}_j(\Delta \bm{\chi}_0(t,s),\{ \bm{\chi}_k(s)\}_{k>0}, \{ \bm{\chi}_k(t)\}_{k>0}) \n\\&-\tilde{\gamma}_j \bm{\chi}_j+\bm{v}_j^{\text{ic}}(\{ \bm{\chi}_k(t)\}) + \bm{\Xi}_j(\{ \bm{\chi}_k(t)\},t),
    \label{eq:effective_dynamics_polymer}
\end{align}
where we denoted by $\Delta \bm{\chi}_0(t,s)=\bm{\chi}_0(t)-\bm{\chi}_0(s)$ the displacement of the rescaled center of mass occurring from time $s$ to time $t\ge s$.
%All terms characterizing the above dynamics are described in what follows.
We now analyze the various terms that appear in Eq.~\eqref{eq:effective_dynamics_polymer}.
First, the interaction of the polymer with the fluctuating field introduces a space- and time-dependent memory kernel $\bm{F}_j$ given by
\begin{align}
    &\bm{F}_j(\Delta \bm{\chi}_0(t,s),\{ \bm{\chi}_k(s)\}_{k>0}, \{ \bm{\chi}_k(t)\}_{k>0})\n \\&=
    D\nu \lambda^2 \int \dslash{q} \,i\bm{q}q^a |V_{\bm{q}}|^2 G_{\bm{q}}(t-s) \sum_{n,k} \varphi_{jk} \mathrm{e}^{ \frac{ i }{\sqrt{N}} \bm{q} \cdot\Delta\bm{\chi}_0(t,s)} \n \\& 
    \quad\times  \exp{- i\bm{q} \cdot\sum_{m>0} 
    [\varphi^{-1}_{nm}\bm{\chi}_m(s)-\varphi^{-1}_{km}\bm{\chi}_m(t)]
    }.
    \label{eq:non_linear_memory_polymer}
\end{align}
Note that the memory kernel depends in a non-linear way on the spatial coordinates~\cite{basu2022dynamics}. Furthermore, its functional dependence on the Rouse modes is qualitatively different, depending on the mode: while the center of mass only appears via its displacement $\Delta \bm{\chi}_0(t,s)$, the higher-order modes $\bm{\chi}_j(s)$ and $\bm{\chi}_j(t)$ at two different times $t$ and $s$ contribute separately to the memory term. 
%The relevance of this apparently innocuous consideration, 
The consequences of this fact, and its effect on the dynamical properties of the polymer, will be clarified below. %in the following derivation.

%Another interesting remark is the fact that 
Interestingly, the memory kernel does not depend on the sign of the interaction coupling, as it is proportional to $\lambda^2\sigma^2=\lambda^2$. This property is a consequence of the original dynamics in Eqs.~\eqref{eq:dynamics_rouse_modes} and~\eqref{eq:field_dynamics_fourier} being invariant under the transformation $(\lambda,\phi) \leftrightarrow (-\lambda,-\phi)$~\cite{venturelli2022nonequilibrium}.
%%%
Importantly, the effective dynamics in Eq.~\eqref{eq:effective_dynamics_polymer} does not rely on any approximation and it is therefore exact to all orders in the interaction coupling $\lambda$.

Next, the statistics of the noise $\bm{\Xi}_j$ in the effective dynamics~\eqref{eq:effective_dynamics_polymer} is also affected by the interaction of the polymer with the order parameter $\phi$. In particular, it consists of a white noise term $\bm{\xi}_j$ which describes the interaction with the thermal bath, and a temporally correlated term %$\bm{\Xi}_j(t)$
resulting from the coarse graining of the field, as shown by the following expression:
\begin{align}
    \bm{\Xi}_j(t)=&\, \bm{\xi}_j(t) + \lambda \sigma \nu \int \dslash{q} i\bm{q} V_{-\bm{q}}  \int_{t_0}^t \dd{s} G_{\bm{q}}(t-s) \zeta_{\bm{q}}(s)  \n \\&\times \sum_{k} \varphi_{jk} \exp[i \bm{q}\cdot \sum_{n}\varphi^{-1}_{kn}\bm{\chi}_n(t) ] .
\label{eq:effective_noise_polymer}
\end{align}
For each wavevector $\bm{q}$, the temporal convolution over the variable $s$ between the response function $G_{\bm{q}}$ and the noise $\zeta_{\bm{q}}$ corresponds to an exponentially colored noise with correlation time $1/\alpha_{\bm{q}}$.
Accordingly, as expected, each mode of the field contributes to the stochastic part of Eq.~\eqref{eq:effective_dynamics_polymer} by introducing some memory over its own typical relaxation time. For this reason, the correlations of $\bm{\Xi}_j$ might extend for arbitrarily long times
%scales 
in the case of locally conserved or critical fields --- see the discussion at the end of Sec.~\ref{sec:model}.  Moreover, the noise $\bm{\Xi}_j$ is multiplicative in that its amplitude depends on the value assumed by the Rouse modes at time $t$.

Finally, at short times, the effective dynamics of the polymer in Eq.~\eqref{eq:effective_dynamics_polymer} is reminiscent of the initial configuration of the field $\phi_{\bm{q}}(t_0)$ via %, as described by 
the following term:
\begin{align}
    \bm{v}_j^{\text{ic}}(t)=&\, \lambda \sigma \nu \int \dslash{q} i\bm{q} V_{-\bm{q}} G_{\bm{q}}(t-t_0) \phi_{\bm{q}}(t_0)  \n \\& \times \sum_{k=0}^{N-1}\varphi_{jk} \exp \left[i \bm{q} \cdot \sum_{n} \varphi^{-1}_{kn} \bm{\chi}_n(t) \right] .
    \label{eq:effective_ic_polymer}
\end{align}
Note that, for an initial field configuration that is spatially constant, the only possibly non-vanishing component $\phi_{\bm{q}}(t_0)$ is the one with $\bm{q}= \bm{0}$, and therefore
$\bm{v}_j^{\text{ic}}(t)$ vanishes at all times.

\subsubsection{Linearized dynamics of the polymer}
\label{subsec:linearized_dynamics}

Unfortunately, not much analytical progress can be made by using directly the effective dynamics in Eq.~\eqref{eq:effective_dynamics_polymer}, due to its non-linear nature and to the complicated statistics of the multiplicative and colored noise $\bm{\Xi}_j(t)$. 
However, since in the following we shall be interested in the long-time asymptotic behavior of the dynamics close to equilibrium, we now focus on the linear response of the system.

Besides, note that linearizing the dynamics may anyhow provide a fairly good approximation even moderately far from 
the equilibrium position.
For example, note that the non-linear dependence of the memory kernel in Eq.~\eqref{eq:non_linear_memory_polymer} on the Rouse modes is weighted by the factor 
$\bm{q}q^a |V_{\bm{q}}|^2 G_{\bm{q}}(t-s)$, 
%$ G_{\bm{q}}(t-s)$, 
which decays exponentially to zero for large wavevectors (see \cref{eq:propagator}). This means that the momentum integral in Eq.~\eqref{eq:non_linear_memory_polymer} has an effective cutoff which depends on the time lag $(t-s)$. For example, in the case of model A, the weight factor is proportional to a Gaussian with standard deviation $\sigma_w=[2 R^2+2D(t-s)]^{-1}$. Consequently, all contributions to the memory kernel coming from momenta $q>q_{\text{cutoff}} \simeq 3 \sigma_w$ are practically negligible. This means that, whenever the displacement of the center of mass $\Delta \bm{\chi}_0(t,s)$ and higher-order Rouse modes at times $t$ and $s$ are much smaller than $1/q_{\text{cutoff}}$, the memory kernel can be linearized around $\Delta \bm{\chi}_0(t,s)=\bm{\chi}_j(s)=\bm{\chi}_j(t)=\bm{0}$.

Accordingly, we linearize the effective dynamics in Eq.~\eqref{eq:effective_dynamics_polymer} and obtain the following approximation for the non-linear memory kernel:
\begin{align}
    &\bm{F}_j(\Delta \bm{\chi}_0(t,s),\{ \bm{\chi}_k(s)\}_{k>0}, \{ \bm{\chi}_k(t)\}_{k>0}) \n\\&\simeq
    D\nu \lambda^2 \int \dslash{q} \,\bm{q}q^a |V_{\bm{q}}|^2 G_{\bm{q}}(t-s)  \n \\& \quad \times  \bm{q} \cdot\sum_{m,n,k=1}^{N-1} \varphi_{jk} [\varphi^{-1}_{nm}\bm{\chi}_m(s)-\varphi^{-1}_{km}\bm{\chi}_m(t)].
    \label{eq:linearizing_F_polymer}
\end{align}
%where all three sums in the last line go from $0$ to $N-1$. 
Furthermore, the last line can be simplified using the properties of the transformation $\bm{\varphi}$, which 
%we remind it 
contains on %along 
its rows the eigenvectors of the connectivity matrix $\bm{M}$ normalized to unity. 
%Specifically,
In fact, since $\bm M$ is symmetric, this implies that $\bm{\varphi}$ is orthogonal, i.e., $\bm{\varphi}^{-1}=\bm{\varphi}^{T}$,
%because $\bm{M}$ is symmetric, 
and that $\varphi_{0j}=1/\sqrt{N}$ for all $j \in \{0,...,N-1 \}$. This can be used to prove the 
simple identity~\cite{doi1988theory}
\begin{equation}
\sum_j\varphi^{-1}_{ji}=\sum_j\varphi_{ij}=\sqrt{N}\sum_j\varphi_{ij}\varphi_{0j}=\sqrt{N}\delta_{i0},
    \label{eq:identity_varphi}
\end{equation}
which can be used to rewrite Eq.~\eqref{eq:linearizing_F_polymer} as 
\begin{align}
    &\bm{F}_j(\Delta \bm{\chi}_0(t,s),\{ \bm{\chi}_k(s)\}_{k>0}, \{ \bm{\chi}_k(t)\}_{k>0}) \n\\&\simeq
    \dot{\Gamma}(t-s) [\bm{\chi}_j(t)-\delta_{j0}\bm{\chi}_0(s)],  
    \label{eq:linearizing_F_polymer_2}
\end{align}
where $\dot{\Gamma}(t)$ is the derivative of the linear memory kernel
\begin{equation}
    \Gamma(t)=\frac{ND\nu \lambda^2}{d} \int \dslash{q} \,  \frac{q^{2+a} |V_{\bm{q}}|^2 G_{\bm{q}}(t)}{\alpha_{\bm{q}}}.
    \label{eq:lin_mem_kernel_polymer}
\end{equation}
Already from Eq.~\eqref{eq:linearizing_F_polymer_2}, one can anticipate that the center of mass and the higher-order Rouse modes behave differently. Indeed, by specializing Eq.~\eqref{eq:linearizing_F_polymer_2} to $j=0$, one can see that the result of the linearization still depends on the displacement $\Delta \bm{\chi}_0(t,s)$, which involves the position of the center of mass at two times $t$ and $s$. Conversely, for $j>0$, Eq.~\eqref{eq:linearizing_F_polymer_2} depends  only  on the Rouse mode $\bm{\chi}_j(t)$ at the single time $t$. Integrating Eq.~\eqref{eq:linearizing_F_polymer_2} over $s$ we get
\begin{align}
    &\int_{t_0}^t \dd{s}  \bm{F}_0(\Delta \bm{\chi}_0(t,s),\{ \bm{\chi}_k(s)\}_{k>0}, \{ \bm{\chi}_k(t)\}_{k>0}) \n\\&\simeq -\int_{t_0}^t ds\, \Gamma(t-s) \dot{\bm{\chi}}_0(s) + \Gamma(t-t_0) \Delta \bm{\chi}_0(t,t_0) 
    \label{eq:linear_memory_polymer}
\end{align}
for the center of mass, and
\begin{align}
    &\int_{t_0}^t \dd{s}  \bm{F}_j(\Delta \bm{\chi}_0(t,s),\{ \bm{\chi}_k(s)\}_{k>0}, \{ \bm{\chi}_k(t)\}_{k>0})\n\\
    &\simeq -[\Gamma(0)-\Gamma(t-t_0)]\bm{\chi}_j(t) 
    \label{eq:linearized_kernel_ho_modes}
\end{align}
for higher-order Rouse modes $\bm{\chi}_j$ with $j>0$.
Equations~\eqref{eq:linear_memory_polymer}~and~\eqref{eq:linearized_kernel_ho_modes} suggest that 
the long-time relaxation of 
%the 
all Rouse
modes will be affected by how fast $\Gamma(t)$ decays over time. The asymptotic behavior of $\Gamma(t)$ at long times turns out to be
\begin{equation}
    \Gamma(t)\sim \begin{cases}
        t^{-d/2-1}\mathrm{e}^{-Drt} & \text{for $r > 0$,}\\
        t^{-d/2} & \text{for $r=0$,}
    \end{cases}
    \label{eq:asymptotics_Gamma_A}
\end{equation}
for model A, and 
\begin{equation}
    \Gamma(t)\sim \begin{cases}
        t^{-d/2-1} & \text{for $r > 0$,}\\
        t^{-d/4} & \text{for $r=0$,}
    \end{cases}
    \label{eq:asymptotics_Gamma_B}
\end{equation}
for model B, as shown in \cref{app:asymptotic_kernel} by inspecting the analytic structure of its Laplace transform 
(in particular, without appealing to a specific form of the interaction potential $V_{\bm{q}}$). 
The scale-free decay of these kernels appears as a manifestation of the slow modes that characterize the medium in model A at criticality ($r=0$), due to the phenomenon of critical slowing down~\cite{hohenberg1977theory,tauber2014critical}, and in model B even off criticality, due to the local conservation of the order parameter $\phi$~\cite{venturelli2022nonequilibrium}.

%To proceed, 
Next, we proceed to the linearization of the noise term $\bm{\Xi}_j(t)$. Keeping only the lowest-order contribution in the Rouse modes in Eq.~\eqref{eq:effective_noise_polymer} leads to the linearized noise $\bm{\Lambda}_j(t)$ defined as
\begin{align}
   \bm{\Lambda}_j(t)&=\bm{\xi}_j(t)      \label{eq:linearized_noise_polymer}\\&
   + \delta_{j0} \sqrt{N}\lambda \sigma \nu \int \dslash{q} i\bm{q} V_{-\bm{q}}  \int_{t_0}^t \dd{s} G_{\bm{q}}(t-s) \zeta_{\bm{q}}(s) . \n
\end{align}
The statistics of
%the linearized noise
$\bm{\Lambda}_j$ is substantially different when acting on the center of mass, i.e., for $j=0$, or on the higher-order Rouse modes $j\ge 1$. %More precisely, 
Indeed,
for the former, the memory effects on the second line of the equation above persist at the level of the linearized dynamics, whereas for the latter they vanish, so that 
%and 
$\bm{\Lambda}_j$ reduces solely to the white noise $\bm{\xi}_j$.
Importantly, it is easy to verify that the stationary time-correlations of $\bm{\Lambda}_0(t)$, which we denote by $C_{\Lambda}^{\alpha \beta}(t-s)$, are related to the linear memory kernel $\Gamma(t-s)$ by the fluctuation-dissipation theorem:
\begin{equation}
    C_{\Lambda}^{\alpha \beta}(t-s) \equiv \langle \Lambda_0^\alpha(t)\Lambda_0^\beta(s) \rangle= \nu T \delta_{\alpha \beta} [2\delta(t-s)+\Gamma(t-s)],
    \label{eq:FDT}
\end{equation}
with $t>s$. 
This feature, which is expected for a system that evolves according to an equilibrium dynamics, is thus correctly reproduced within the linearized theory.

Finally, one can linearize the contribution $\bm{v}^{\text{ic}}_0$ in Eq.~\eqref{eq:effective_ic_polymer} around $\{\bm{\chi}_j(t)=0\}$, finding, for a generic initial configuration of the field $\phi_{\bm{q}}(t_0)$,
\begin{align}
    \bar{\bm{v}}_j^{\text{ic}}(t)=&\,\lambda \sigma \nu \int \dslash{q} i\bm{q} V_{-\bm{q}} G_{\bm{q}}(t-t_0) \phi_{\bm{q}}(t_0)  
    \n \\& \times 
    \left[\sqrt{N}\delta_{j0} + i \bm{q}\cdot \bm{\chi}_j(t) \right].
    \label{eq:linearized_ic_polymer}
\end{align}
By using 
\cref{eq:linear_memory_polymer,eq:linearized_noise_polymer,eq:linearized_ic_polymer},
the effective linearized dynamics of the center of mass takes the form of a 
%generalized Langevin equation 
GLE and is given by
\begin{align}
   \dot{\bm{\chi}}_0(t)=&-\int_{t_0}^t ds\,\Gamma(t-s)\dot{\bm{\chi}}_0(s) - [\tilde{\gamma}_0-\Gamma(t-t_0)]\bm{\chi}_0(t)\n\\
   &-\Gamma(t-t_0)\bm{\chi}_0(t_0) +\bm{\Lambda}_0(t)+\bar{\bm{v}}_0^{\text{ic}}(t).
    \label{eq:GLE_com_polymer_GENERAL}
\end{align}
Besides the non-Markovian nature of this effective dynamics, we note that marginalizing the field introduces two effects. First, it modifies the typical (inverse) relaxation time $\tilde{\gamma}_0$ associated with the harmonic confining potential. Second, it adds a time-dependent force proportional to the initial center-of-mass position $\bm{\chi}_0(t_0)$, which decays over time in the same way as the linear memory kernel $\Gamma(t-t_0)$ in Eq.~\eqref{eq:lin_mem_kernel_polymer}.

For the case of the higher-order Rouse modes $\bm{\chi}_j$ with $j>0$, the effective linearized dynamics
obtained from \cref{eq:linearized_kernel_ho_modes,eq:linearized_noise_polymer,eq:linearized_ic_polymer}
reads, instead,
\begin{align}
   \dot{\bm{\chi}}_j(t)=&- [\tilde{\gamma}_j+\Gamma(0)]\bm{\chi}_j(t)+\Gamma(t-t_0)\bm{\chi}_j(t) \n\\
   &+\bm{\xi}_j(t)+\bar{\bm{v}}_j^{\text{ic}}(t).
    \label{eq:eff_lin_dyn_ho_modes}
\end{align}
As anticipated, this resulting dynamics is actually Markovian, because the memory effects introduced by integrating out the fluctuating order parameter $\phi$ did not survive the linearization procedure. Also in this case, the marginalization of the field introduces a time-dependent 
(positive, see \cref{eq:lin_mem_kernel_polymer}) 
correction to the relaxation rate $\tilde{\gamma}_j$. At variance with the case of $\bm{\chi}_0$ in Eq.~\eqref{eq:GLE_com_polymer_GENERAL}, this correction does not actually vanish at long times but it reaches the (positive) value of $\Gamma(0)$. 
Interestingly, all Rouse modes are affected by the same correction.

%%
%%
%%
%%%%%%%%%%%%%%%%%%%%%%%%%%%%%%%%%%%%%%%%%%%%%%%%%%%%%%%%%
%%%%%%%%%%%%%%%%%%%%%%%%%%%%%%%%%%%%%%%%%%%%%%%%%%%%%%%%%
\begin{figure}[t]
    \includegraphics[width=0.9\linewidth]{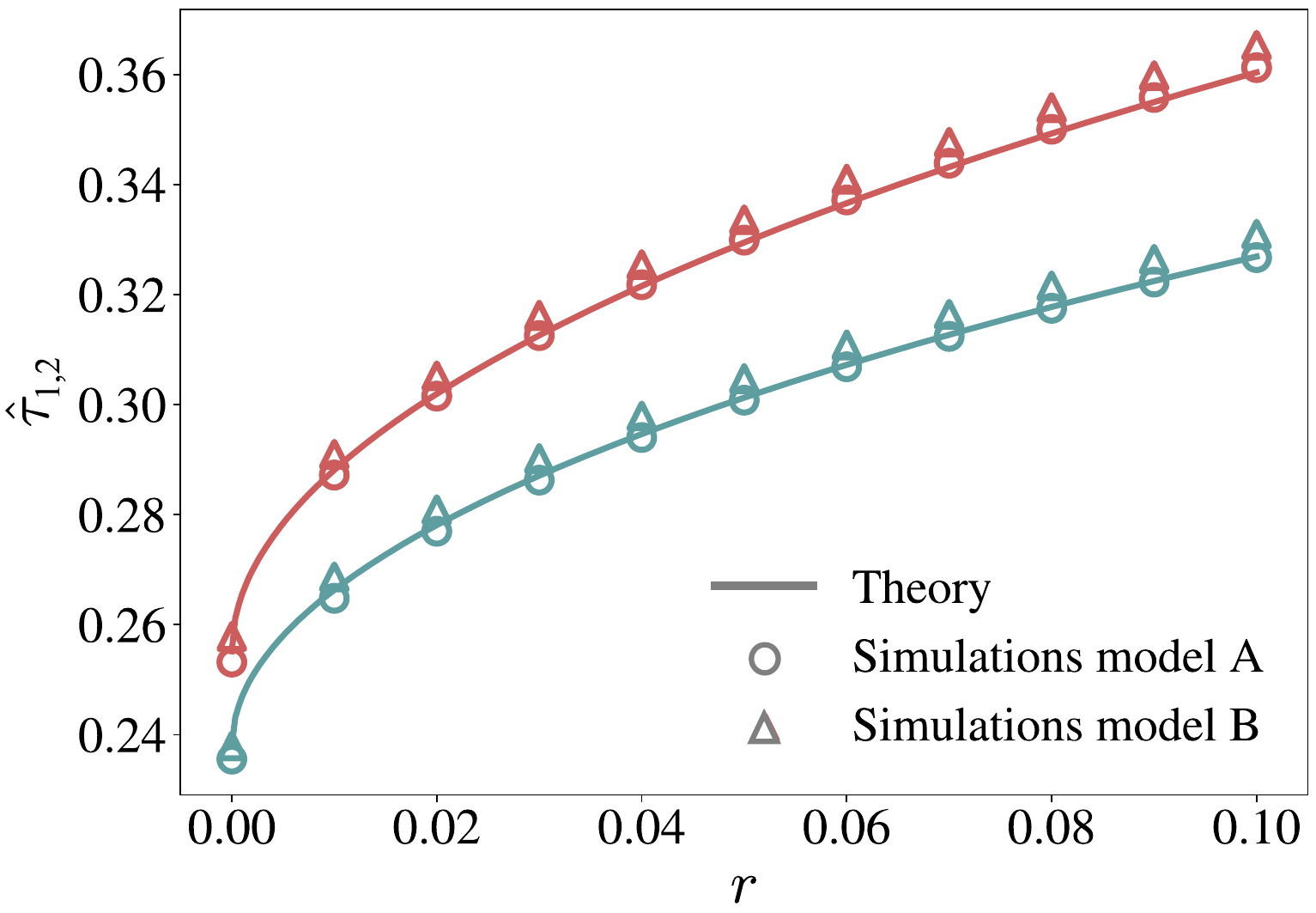}
    \caption{Relaxation time $\hat{\tau}_{1,2}$ of the Rouse modes $\bm{\chi}_1$ and $\bm{\chi}_2$ as functions of the deviation $r$ from criticality. The plot shows that the coupling with a fluctuating and correlated medium speeds up the relaxation of the internal structure of the polymer. This effect is more pronounced when the field is characterized by correlations on large length scales (i.e., as $r \to 0$). The theoretical prediction (solid line) $\hat{\tau}_j \equiv 1/[\tilde{\gamma}_j+\Gamma(0)]$ has been obtained on the basis of Eqs.~\eqref{eq:eff_lin_dyn_ho_modes}, \eqref{eq:relaxation_rates_polymer}, and~\eqref{eq:lin_mem_kernel_polymer}. The simulation results (symbols), instead, were obtained by extracting the slopes  $1/\hat{\tau}_{1,2}$ of the relaxation of the Rouse modes (obtained via molecular dynamics simulations) in logarithmic scale via a linear fit \rev{(see Fig.~\ref{fig:linear_fitting})}.
    %in App.~\ref{sec:simulations})}.  
    The two lines tend to their respective asymptote $1/\tilde{\gamma}_{1,2}$. This plot was obtained with a \rev{linear} polymer of $N=10$ monomers, while all other parameters were set to unity.
    }
    \label{fig:relaxation_ho_modes}
\end{figure} 
%%%%%%%%%%%%%%%%%%%%%%%%%%%%%%%%%%%%%%%%%%%%%%%%%%%%%%%%%
%%%%%%%%%%%%%%%%%%%%%%%%%%%%%%%%%%%%%%%%%%%%%%%%%%%%%%%%%
%%
%%
%%

\begin{figure*}[t]
    %\centering
    %\begin{subfigure}{0.49\linewidth}
    \subfloat{
        \includegraphics
        %[width=\linewidth, height=6.2cm]
        [width=0.49\linewidth]
        {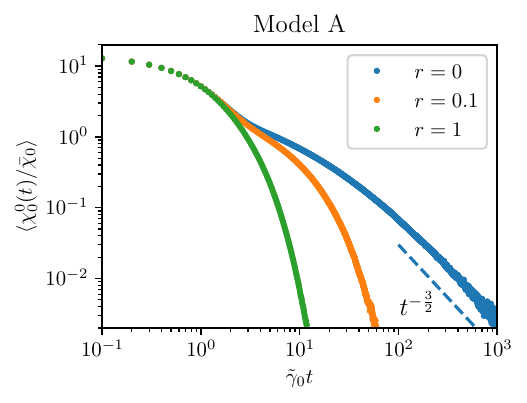}
        \label{3a}
    }
    %\end{subfigure}
    %\hfill
    %\begin{subfigure}{0.49\linewidth}
    \subfloat{
        \includegraphics
        %[width=\linewidth, height=6.2cm]
        [width=0.49\linewidth]
        {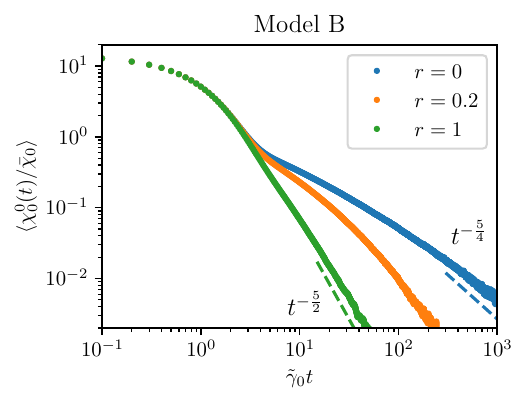}
        \label{3b}
    }
    %\end{subfloat}
    \caption{Nonequilibrium relaxation of the average center of mass of a \rev{linear} polymer initially displaced from the rest position of the trap by an amount $\bar{\chi}_0$, in spatial dimension $d=1$. 
    The symbols indicate the result of numerical simulations (see App.~\ref{sec:simulations} for details), with the field initialized in the flat configuration $\phi_{\bm{q}}=0$ for all momenta $\bm{q}$. As specified in the legend, each color corresponds to a different correlation lengths $\xi_\phi=r^{-1/2}$ of the field. Left and right panels report the relaxation of $\langle \chi_0 \rangle$ obtained with the field evolving according to model A or model B dynamics, respectively. 
    The algebraic decay of this quantity at long times, theoretically predicted in \cref{eq:relax_modelA,eq:relax_modelB}, is indicated by the dashed lines. 
    The plots show that the relaxation of the center of mass towards the bottom of the confining potential is slowed down by the order parameter field, especially when the latter is critical.
    In the simulation we used $N=10$ and $T=0.01$, while all other parameters were set to unity.
    } 
    \label{fig:relaxation_com}
\end{figure*}
%%%%%%%%%%%%%%%%%%%%%%%%%%%%%%%%%%%%%%%%%%%%%%%%%%%%%%%%%
%%%%%%%%%%%%%%%%%%%%%%%%%%%%%%%%%%%%%%%%%%%%%%%%%%%%%%%%%
%%
%%
%%

In particular, when all the monomers have the same coupling with the medium, the attractive field-mediated forces cause a speed-up of the relaxation of the polymer internal structure, which nonetheless remains exponential 
(as in the absence of interactions with the medium). 
To analyze this effect,
in Fig.~\ref{fig:relaxation_ho_modes}
we report the relaxation times of the Rouse modes $\bm{\chi}_1$ and $\bm{\chi}_2$ as functions of the deviation $r$ from criticality, as measured in molecular dynamics simulations (symbols\rev{, see App.~\ref{sec:simulations} for the details}), and as predicted by Eqs.~\eqref{eq:eff_lin_dyn_ho_modes} and~\eqref{eq:lin_mem_kernel_polymer} in the linear-response regime (solid lines). The agreement is excellent not only for the specific choice of parameters used in the figure. 
Note that, as predicted by Eq.~\eqref{eq:lin_mem_kernel_polymer}, taking into account Eq.~\eqref{eq:def-alphaq}, the relaxation times of the Rouse modes $\bm{\chi}_i$ for $i\ge 1$ is actually independent of the dynamics of the field being conserved or not.

\subsubsection{Long-time relaxation of the center of mass}

We now analyze the non-equilibrium relaxation of the center of mass of the polymer towards the bottom of the confining potential, after an initial displacement $\bm{\chi}_0(t_0)$. 
The analogous problem was investigated in Ref.~\cite{venturelli2022nonequilibrium} for a single particle coupled to a fluctuating Gaussian field. In this case, after an initial rearrangement in the neighborhood of the particle, the field lags behind the moving particle and it produces a slowing down of the relaxation process. Based on this heuristic picture, we thus expect a similar slowing effect in the dynamics of the polymer, as we confirm below.

Specifically, without loss of generality, we set
%that 
$t_0=0$ and assume that the center of mass of the polymer is initially displaced from the center of the trap (where the corresponding force vanishes) by an amount $\bar{\chi}_0$ along the $\alpha$-direction, i.e., $\chi_0^\beta(t_0)=\bar{\chi}_0\delta_{\alpha \beta}$. 
The resulting dynamics can be conveniently analyzed in the Laplace domain by first solving for $\hat{\chi}^\alpha_0(z)=\int_0^\infty \dd t\,\expval{ \chi^\alpha_0(t)}\mathrm{e}^{-zt}$, i.e., the Laplace transform of the average position of the center of mass along the $\alpha$-direction, and then studying its analytic structure in the complex $z$ plane~\cite{venturelli2023memory,hull1955asymptotic}.
Unfortunately, \cref{eq:GLE_com_polymer_GENERAL} cannot be immediately solved by taking its Laplace transform because of the presence of the terms $\Gamma(t-t_0)\bm{\chi}_0(t)$ and $\bar{\bm{v}}_0^{\text{ic}}(t)$, which involve products of $t$-dependent functions. However, since the initial condition of the field is not expected to influence the long-time behavior of the system~\footnote{Further comments on this %point
are presented in \cref{app:asymptotic}.},
%~\cite{venturelli2022nonequilibrium}, 
we make the convenient choice 
\begin{equation}
\phi_{\bm{q}}(t_0)=ND\lambda \sigma V_{\bm{q}}q^a /\alpha_{\bm{q}}.
    \label{eq:initial_condition_field}
\end{equation}
With this initial condition, the linearization of Eq.~\eqref{eq:effective_ic_polymer} can be shown to give $\bm{v}_0^{\text{ic}}(t)=-\Gamma(t-t_0)\bm{\chi}_0(t)$, which simplifies the GLE as
\begin{align}
    \dot{\bm{\chi}}_0(t)=&-\int_{t_0}^t ds\,\Gamma(t-s)\dot{\bm{\chi}}_0(s) - \tilde{\gamma}_0\bm{\chi}_0(t)\n\\&-\Gamma(t-t_0)\bm{\chi}_0(t_0) +\bm{\Lambda}_0(t).
    \label{eq:GLE_com_dimer}
\end{align}
The average value of the solution of \cref{eq:GLE_com_dimer} can now be expressed as
\begin{equation}
%\hat{\chi}^\alpha_0(z)=\frac{\bar{\chi}_0}{\tilde{\gamma}_0+z+z\hat{\Gamma}(z)},
   %\ag{better:}
   \hat{\chi}^\beta_0(z)=\frac{\bar{\chi}_0}{\tilde{\gamma}_0+z+z\hat{\Gamma}(z)}\delta_{\alpha\beta},
   \label{eq:GLE_com_dimer_laplace}
\end{equation}
where $\hat \Gamma(z)$ is the Laplace transform of $\Gamma (t)$. This shows (as expected) that the asymptotic behavior of the average position of the center of mass is strictly related to that of the linear memory kernel $\Gamma(t)$,
which we studied above in \cref{eq:asymptotics_Gamma_A,eq:asymptotics_Gamma_B}. 
In particular, in \cref{app:asymptotic_com}, on the basis of Eq.~\eqref{eq:GLE_com_dimer_laplace} we determine the following asymptotic behaviors:
\begin{equation}
    \expval{ \chi^\alpha_0(t)} \sim
    \begin{cases}
        t^{-\left(2+d/2 \right)} & \T{for}  \quad   r>0, \\
        t^{-\left(1+d/4 \right)} & \T{for}  \quad  r=0,
    \end{cases}
    \label{eq:relax_modelB}
\end{equation}
for model B, while
\begin{equation}
    \expval{ \chi^\alpha_0(t)} \sim
                       t^{-\left(1+d/2\right)} \mathrm{e}^{-Dr t} 
    \label{eq:relax_modelA}
\end{equation}
for model A with $r\ll \tilde\gamma_0/D$, whereas the decay becomes purely exponential for $r\gg \tilde\gamma_0/D$ (see \cref{app:asymptotic_com} for details).
The position of the center of mass thus exhibits an algebraic decay for model B, with an exponent that depends on whether the medium is critical ($r=0$) or not ($r>0$). 
As mentioned above,
this originates from the local conservation law that characterizes the evolution of the medium in model B dynamics, thus producing slow modes $\bm q\sim 0$~\cite{hohenberg1977theory,tauber2014critical}. By contrast, for model A, an algebraic decay is only found when $r=0$, since in this case the relevant mechanism is the critical slowing down that affects the medium when $r=0$~\cite{hohenberg1977theory,tauber2014critical,venturelli2022nonequilibrium}. 
Figure~\ref{fig:relaxation_com} shows, in a double logarithmic scale, the evolution of $\langle 
\bm{\chi}_0(t)\rangle$ along the direction of the initial displacement $\bar\chi_0$ (taken to be $\alpha=0$, corresponding to the $x$-axis), 
for model A (left) or model B (right) dynamics and various values of the distance $r$ from the critical point. The behavior at long times predicted in \cref{fig:relaxation_com} and indicated by the dashed lines in both panels is compared with the results of numerical simulations (performed as described in App.~\ref{sec:simulations}), finding a very good agreement. 

\section{Typical polymer size}
\label{sec:polymer_size}

In this Section we study the average square
gyration radius $\langle R_g^2 \rangle$ and end-to-end distance $\langle \bm{R}_{\text{ee}} \rangle$ of the polymer in the steady state, when its internal structure has already relaxed to an equilibrium configuration. Both quantities are a measure of the typical polymer size, and are defined as
\begin{align}
    R_g^2 &=\frac{1}{N}\sum_{n=0}^{N-1} (\bm{X}_n-\bm{X}_{\text{com}})^2=\frac{1}{N} \sum_{n \neq 0} \bm{\chi}^2_n, \label{eq:def_Rg}\\
    \bm{R}_{\text{ee}}&=\bm{X}_{N-1}-\bm{X}_{0}=\sum_{n \neq 0} (\varphi_{n,N-1}-\varphi_{n,0}) \bm{\chi}_n. \label{eq:def_Ree}
\end{align}
However, while the 
%first one 
former
is appropriate for any topology of the chain, i.e., for any connectivity matrix $\bm{M}$, the 
%second one 
latter
is only meaningful for a linear chain, where the two ends can be clearly identified. 
%notion of the two ends is better defined. 
Although the derivation would be analogous for a generic topology of the polymer, in the following we focus on a linear chain with terminal monomers given by $\bm{X}_0$ and $\bm{X}_{N-1}$.

To keep the derivation more general, we also consider the possible presence of a stretching force $\bm{f}_s$ that acts on the
terminal monomers, hence on the
end-to-end distance $\bm{R}_{\text{ee}}$. In particular, we add the 
%following 
stretching potential
\begin{equation}
    \mathcal{U}_f=-\bm{f}_s \cdot \bm{R}_{\text{ee}}=-\bm{f}_s \cdot \sum_i (\varphi_{i,N-1}-\varphi_{i,0}) \bm{\chi}_i
    \label{eq:stretching_potential}
\end{equation}
to the total Hamiltonian $\mathcal{H}_0 + \mathcal{H}_{\text{eff}}$ reported in Eq.~\eqref{eq:marginal_distribution_polymer}. The response of the polymer to such a stretching force, described by the so-called force-extension curve, 
%is analyzed in a dedicated subsection.
will be analyzed in \cref{sec:force-extension}.
In general, both $\langle R_g^2 \rangle$ and $\langle \bm{R}_{\text{ee}} \rangle$ are affected by the fluctuating order parameter $\phi$ of the medium. The effect of the field-mediated forces on the typical size of the polymer is studied in 
%the following 
\cref{sec:typical_no_f}
%with the help of 
via a perturbative expansion in the interaction coupling $\lambda$,
developed in \cref{sec:weak-coupling-size}.
Whenever this is possible, we compare the results of this weak-coupling approximation with the theoretical predictions obtained from the linearized theory derived in Section~\ref{subsec:linearized_dynamics}.
Finally, in \cref{sec:moving-trap} we shall assume that the harmonic potential acting on 
the monomers is displaced in space at constant velocity $\bm v$, driving the system in a non-equilibrium stationary state.  This allows us to probe, as a function of $v=|\bm v|$, how the typical size of the polymer is modified by the presence of nonequilibrium field-mediated forces.

\subsection{Weak-coupling approximation}
\label{sec:weak-coupling-size}

As in Section \ref{sec:relaxation}, we consider here the case in which all monomers have the same coupling with the field, i.e., $\sigma_i=\sigma$ independently of $i$, and thus that they experience an effective attraction. In the framework of the weak-coupling approach, we find it convenient to introduce the average over the stationary distribution of the Rouse modes
%$\langle \bullet \rangle_{f,\lambda}$,
\begin{equation}
    \langle \dots \rangle_{f,\lambda} = \frac{1}{\mathcal{N}} \prod_{j=0}^{N-1} \int \dd \bm{\chi}_j \, (\dots)\, \mathrm{e}^{ -\beta (\mathcal{H}_0 + \mathcal{H}_{\text{eff}} +\cor U_f )},
    \label{eq:stat_average}
\end{equation}
where the modulus $f$ of the stretching force and the magnitude $\lambda$ of the interaction coupling with the field are explicitly indicated as subscripts. 
Recall that the effective interaction Hamiltonian $\cor H_\T{eff}$, proportional to $\lambda^2$, was introduced in \cref{eq:eff_Hamiltonian_polymer}, while the prefactor $\cor N^{-1}$ in \cref{eq:stat_average} ensures that $\expval{1}_{f,\lambda}=1$.
In particular, we want to obtain the first non-trivial correction to $\langle R_g^2 \rangle_{f,0}$ and $\langle \bm{R}_{\text{ee}} \rangle_{f,0}$, which is induced by the coupling of the polymer with the field. Using a standard perturbative expansion in $\lambda$, we get
\begin{align}
    \langle O \rangle_{f,\lambda}-\langle O \rangle_{f,0}=&-\beta \left[ \langle O \mathcal{H}_{\text{eff}}\rangle_{f,0} - \langle O \rangle_{f,0}  \langle \mathcal{H}_{\text{eff}} \rangle_{f,0}  \right] \n\\& + \mathcal{O}(\lambda^4) \label{eq:pert_exp},
\end{align}
where the observable $O$ can be replaced by either $R_g^2$ or $\bm{R}_{\text{ee}}$. Note that the correction on the r.h.s.~of the first line 
involves only averages computed over the uncoupled system (i.e., with $\lambda=0$); yet, since $\mathcal H_\mathrm{eff}\propto\lambda^2$, this correction will be shown to be of $\order{\lambda^2}$.
In particular, all corrections proportional to odd powers of $\lambda$ should vanish, due to the symmetry (mentioned above) of the equations of motion~\eqref{eq:dynamics_rouse_modes} and~\eqref{eq:field_dynamics_fourier} under $(\lambda,\phi) \leftrightarrow (-\lambda,-\phi)$~\cite{venturelli2022nonequilibrium,basu2022dynamics}.

All the averages needed to evaluate Eq.~\eqref{eq:pert_exp} can be computed with the help of the generating functional $\mathcal{Z}[\{\vb j_i\}]$ of the free Rouse chain, which can be constructed as
%is defined as
\begin{align}
    &\mathcal{Z}[\{\vb j_i\}]= \Big \langle \exp \Big(\sum_{i} \vb j_i \cdot \bm{\chi}_i  \Big) \Big \rangle_{f,0} \label{eq:generating_functional}\\&=
    \exp{ \frac{1}{2\beta} \sum_i \frac{1}{\mathcal{M}_i} [\vb j_i^2 + 2 \beta (\varphi_{i,N-1}-\varphi_{i,0})\bm{f}_s\cdot \vb j_i]} \n,
\end{align}
where we introduced the quantity $\mathcal{M}_i=\kappa m_i + \kappa_c$, with $m_i$ the eigenvalues of the connectivity matrix $\bm M$. The expression of $\mathcal{Z}[\{\vb j_i\}]$ given in Eq.~\eqref{eq:generating_functional} can be obtained by the standard methods discussed in~\cref{app:corr_Rg}. In particular, the generating functional can be readily used to derive the expressions of the unperturbed $\langle R_g^2 \rangle_{f,0}$ and $\langle \bm{R}_{\text{ee}} \rangle_{f,0}$. These are, respectively, given by
\begin{align}
&\langle R_g^2 \rangle_{f,0} =\frac{1}{N} \sum_{n \neq 0} \langle \bm{\chi}^2_n \rangle_{f,0}= \frac{1}{N} \sum_{n \neq 0} \sum_{\alpha=0}^{d-1} \eval{\frac{\partial^2 \mathcal{Z}[\{ \vb j_i\}]}{\partial j_n^\alpha \partial j_n^\alpha}}_{\vb j_i=\bm{0}} \n\\& =\frac{1}{N} \sum_{n \neq 0} \left\lbrace \frac{d}{\beta \mathcal{M}_n} + \left[ \frac{(\varphi_{n,N-1} - \varphi_{n,0}) \bm{f}_s}{\mathcal{M}_n}\right]^2\right\rbrace,
\label{eq:unperturbed_Rg2}
\end{align}
and
\begin{align}
    &\langle R^\alpha_{\text{ee}} \rangle_{f,0}=\sum_n(\varphi_{n,N-1}-\varphi_{n,0}) \eval{\frac{\partial \mathcal Z[\{ \vb j_i\}]}{\partial j_n^\alpha }}_{\vb j_i = \bm 0} \n\\&= \sum_n \frac{(\varphi_{n,N-1}-\varphi_{n,0})^2}{\mathcal{M}_n}f\delta_{\alpha0},
    \label{eq:unperturbed_Ree}
\end{align}
where, without loss of generality, we assumed that the stretching force is directed along the $x$-axis, so that $\bm{f}_s=f\hat{\bm{e}}_0$. 
The correction of $\mathcal O(\lambda^2)$ to the unperturbed values in Eqs.~\eqref{eq:unperturbed_Rg2} and~\eqref{eq:unperturbed_Ree}, induced by the coupling of the polymer with the field,
%depends 
turns out to depend on averages of the type $\langle \bm{\chi}^p_j \exp [i \bm{q} \cdot (\bm{X}_k-\bm{X}_n)]\rangle$, with $j$, $k$, and $n$ generic indices in the set $\{ 0,1,...,N-1\}$, and the power $p \in \{ 0,1,2\}$. These averages can be computed again with the help of the generating functional $\mathcal{Z}[\{\vb j_i\}]$ as detailed in~\cref{app:corr_Rg}. This way we obtain
\begin{widetext}
\begin{align}
    \expval{R_g^2}_{f,\lambda} -\expval{R_g^2}_{f,0}
    =&\, \frac{\lambda^2 \beta}{2N} \sum_{n \neq 0} \sum_{ij}\int \dslash{q}  \abs{V_q}^2 \cor C\q \left[\frac{2 i \beta^{-1} \bm{f}_s\cdot \bm{q} \left(\varphi_{n, N-1}-\varphi_{n, 0}\right)\left(\varphi_{ni}-\varphi_{nj}\right) -q^2\beta^{-2}\left(\varphi_{ni}-\varphi_{nj}\right)^2}{\mathcal M_n^2}\right] \nonumber \\
    &\times \exp[-\frac{q^2}{2 \beta} \sum_l \frac{(\varphi_{li}-\varphi_{lj})^2}{\mathcal M_l}+i \bm{f}_s \cdot \vb q \sum_l  \frac{(\varphi_{li}-\varphi_{lj})\left(\varphi_{l, N-1}-\varphi_{l, 0}\right)}{\mathcal M_l}  ] + \mathcal{O}(\lambda^4), 
    \label{eq:Rg_correction}
\end{align}
and
\begin{align}
        \langle R^\alpha_{ee} \rangle_{f,\lambda}-\langle R^\alpha_{ee} \rangle_{f,0}=&\, \frac{ \lambda^2 \beta}{2}   \sum_{n \neq 0}\sum_{ij}\int \dslash{q}   \frac{\abs{V_q}^2 \cor C\q}{\mathcal{M}_n}\left[ i q^\alpha \beta^{-1} (\varphi_{ni}-\varphi_{nj})(\varphi_{n,N-1}-\varphi_{n,0})\right] \n \\&
        \times \exp[-\frac{q^2}{2 \beta} \sum_l \frac{(\varphi_{li}-\varphi_{lj})^2}{\mathcal M_l}+i \bm{f}_s \cdot \vb q \sum_l  \frac{(\varphi_{li}-\varphi_{lj})\left(\varphi_{l, N-1}-\varphi_{l, 0}\right)}{\mathcal M_l}  ] + \order{\lambda^4}.\label{eq:Re_correction}
\end{align}
\end{widetext}

In the next two Sections, we analyze these expressions separately in the cases $f=0$ and $f\neq 0$.

%%
%%
%%
%%%%%%%%%%%%%%%%%%%%%%%%%%%%%%%%%%%%%%%%%
%%%%%%%%%%%%%%%%%%%%%%%%%%%%%%%%%%%%%%%%%
\begin{figure*}[t]
    %\centering
    %\begin{subfigure}{0.49\linewidth}
    \subfloat{
        \includegraphics
        %[width=\linewidth, height=6.2cm]
        [width=0.49\linewidth,height=6.2cm]
        {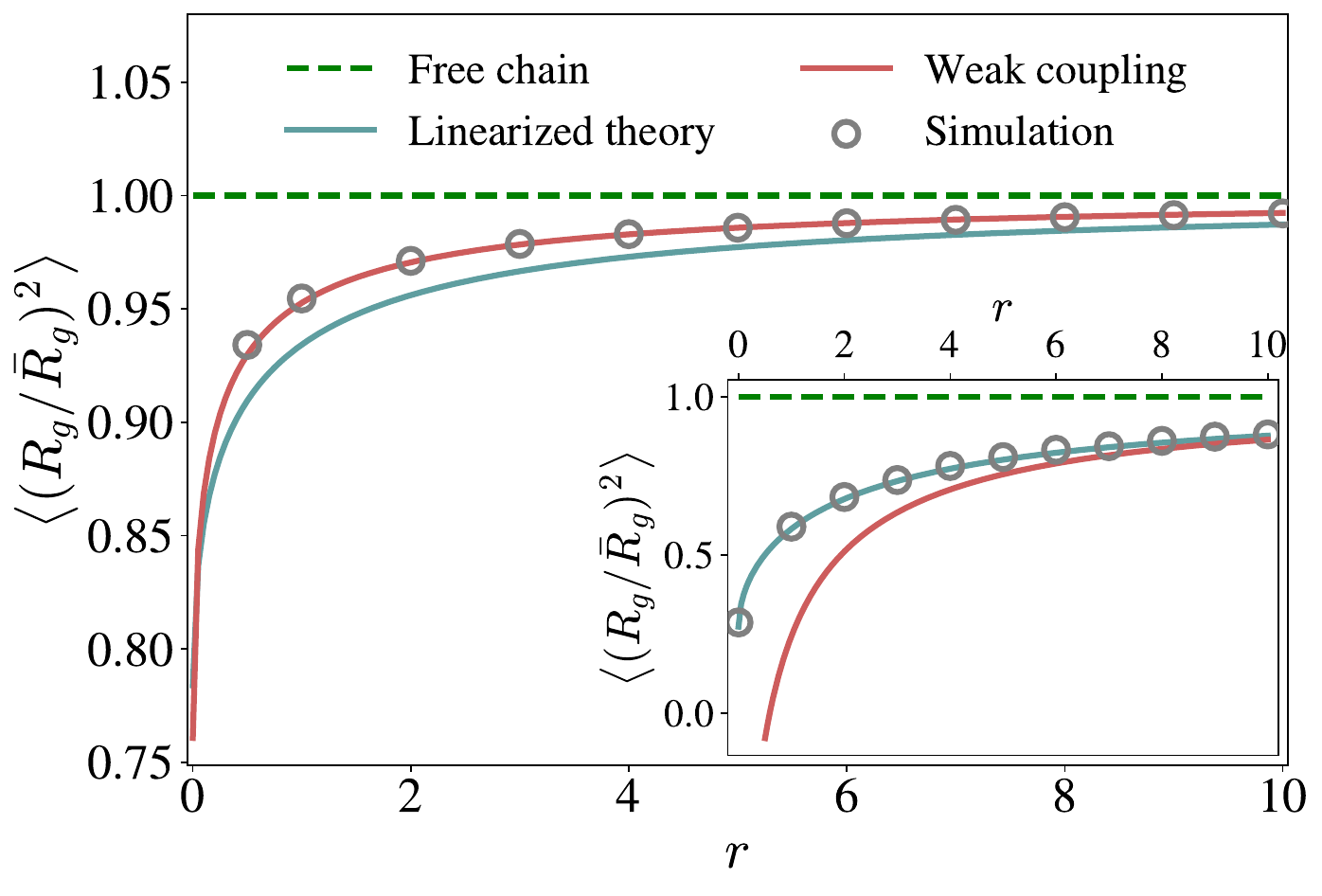}
        \label{fig:4a}
    }
    %\end{subfigure}
    %\hfill
    %\begin{subfigure}{0.49\linewidth}
    \subfloat{
        \includegraphics
        %[width=\linewidth, height=6.2cm]
        [width=0.49\linewidth, height=6.2cm]
        {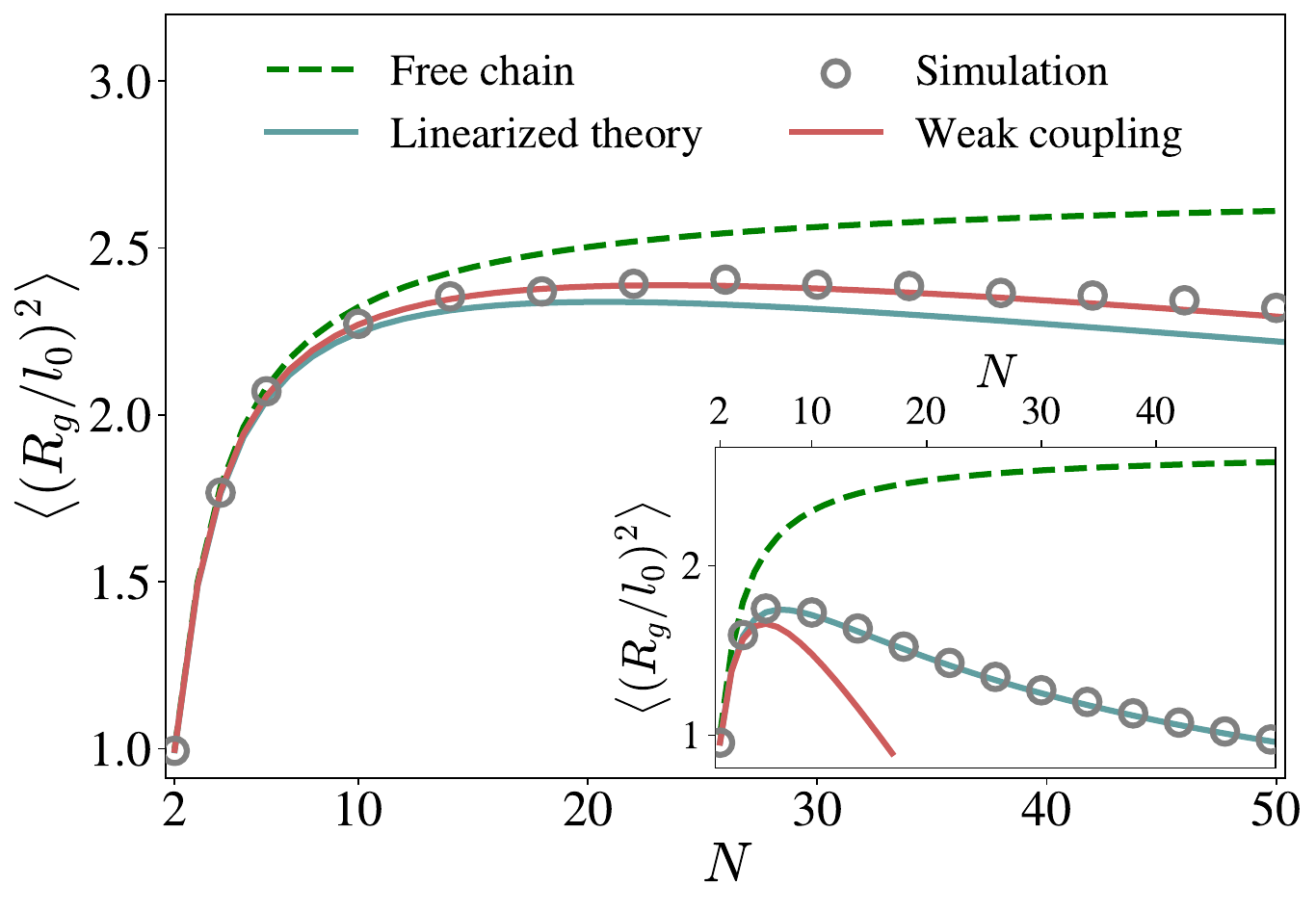}
        \label{fig:4b}
    }
    %\end{subfloat}
    \caption{Typical size of \rev{a linear} polymer at equilibrium, quantified by the mean-square gyration radius $\langle R_g^2\rangle$ (see Eq.~\eqref{eq:def_Rg}) 
    in the absence of stretching forces. The results of the numerical simulations in $d=1$ (grey symbols) are compared to the theoretical predictions obtained with either the weak-coupling approximation (red lines, see \cref{eq:Rg_correction}), or the linearized theory (light blue lines, see \cref{eq:Rg_linearized_theory}).  
    The choice of parameters in the main plots is such that the weak-coupling approximation 
    %produces better results than the 
    is more accurate than the
    linearized theory. 
    The two insets show instead that, with higher coupling $\lambda$ and lower temperature $T$, the linearized theory is more accurate.
    Left panel: $\langle R_g^2\rangle$ of a linear chain with polymerization degree $N=20$ as a function of $r=1/\xi^2_{\phi}$. 
    The value of $\langle R_g^2\rangle$ is measured in units of $\bar{R}_g^2$, i.e., of the gyration radius of a free chain 
    (i.e., in the absence of interaction with the medium)
    with the same parameters (green dashed line). 
    This value is given in \cref{eq:unperturbed_Rg2} upon setting $\bm{f}_s=\bm 0$.
    The figure shows that, when the field approaches the critical point $r=0$, the typical polymer size is reduced as a consequence of the larger field-mediated forces. 
    %Simulation parameters in the main plot: $\kappa=1$, $\nu=1$, $T=1$, $D=1$, $R=1$, $\lambda=0.3$. In the inset: $\lambda=1$ and $T=0.1$. 
    In the main plot we used $\lambda=0.3$ and $T=1$, while in the inset we chose $\lambda=1$ and $T=0.1$ (all other parameters were set to unity in both cases).
    Right panel: $\langle R_g^2\rangle$ of a linear chain as a function of $N$, measured in units of the bond length $l_0=dT/(\kappa + \kappa_c)$. 
    For a sufficiently large $N$, the field-mediated forces induce a collapse of the chain, as shown by the non-monotonic behavior of the curves. The simulation parameters are the same as in the left panel, with $r=1$.}
    \label{fig:Rg}
\end{figure*} 
%%%%%%%%%%%%%%%%%%%%%%%%%%%%%%%%%%%%%%%%%

\subsection{Typical size in the absence of external forces}
\label{sec:typical_no_f}

%It
First, it is straightforward to verify that the correction to the end-to-end distance $\bm{R}_{\text{ee}}$ due to the interaction with the field vanishes when $\bm{f}_s=\bm{0}$, as expected from simple symmetry arguments. 
In particular, for any given polymer configuration $\{\bm{X}_i\}$, the one obtained by
the transformation $\{\bm{X}_i\} \to \{-\bm{X}_i\}$ has the same statistical weight but opposite
end-to-end distance.
In this case, when no stretching force is applied to the polymer, we focus on Eq.~\eqref{eq:Rg_correction} to study how the polymer size is affected by the interaction with the  correlated medium.

The analytical result in Eq.~\eqref{eq:Rg_correction} is shown in Fig.~\ref{fig:Rg}. In particular, we plot $\langle R_g^2 \rangle_{0,\lambda}$ as a function of the distance $r$ from the critical point (left panel) or of $N$, for a fixed value of $r$ (right panel). 
In the first case, we observe that the gyration radius of the polymer decreases when the medium develops long-range spatial correlations, i.e., upon decreasing $r$. This 
%has to do with 
is because the forces induced by the near-critical field are significant over an increasingly larger length scale. 
Importantly, unlike previous studies of polymeric macromolecules dispersed in near-critical binary liquid mixtures~\cite{DeGennes1976, deGennes_1980, vilgis1993conformation}, our theoretical model does not predict a re-expansion of the polymer when the field reaches the critical point. 
This %The absence of this effect 
%has to do 
is consistent
with the fact that the field $\phi$ is %being 
described by a Gaussian theory \rev{(which is unable to account for the emergence of ``droplets'' below the critical point, i.e., for regions with $\expval{\phi}\neq 0$)}, 
as pointed out in Ref.~\cite{vilgis1993conformation}. 
The right panel of Fig.~\ref{fig:Rg} shows the dependence of the typical size of the polymer on 
the polymerization degree $N$ at a fixed distance $r$ from the critical point.
The value of the gyration radius is measured in units of the typical length scale $l_0=dT/(\kappa + \kappa_c)$, which is of the order of the bond size in the absence of the field (see, e.g., Ref.~\cite{doi1988theory}).
For sufficiently short chains, we observe that adding a monomer increases $\langle R_g^2 \rangle_{0,\lambda}$.
This expected behavior has an entropic origin, and it is quantitatively similar to the one observed for a chain decoupled from the field, i.e., with $\lambda=0$ (dashed green line). However, after some threshold value of $N$, the pairwise-additive field-mediated forces dominate the entropic effect, resulting into a decrease of $\langle R_g^2 \rangle_{0,\lambda}$ upon increasing the number $N$ of monomers. 
In the case of an ideal chain as the one considered here, i.e., in the absence of excluded-volume interactions and steric hindrance effects, the size of the polymer can become arbitrarily small as $N$ increases.
In a realistic chain, instead, the polymer would first collapse into a dense compact globule, and then its size would increase as $\sim N^{1/3}$ in $d=3$ --- as in the case of a polymer chain in a poor solvent, where the contacts with the solvent molecules are minimized~\cite{rubinstein2003polymer, degennes1979scaling}. 

In both panels of Fig.~\ref{fig:Rg}, the theoretical predictions obtained within the weak-coupling approximation (solid red lines) are compared with those derived from the linearized theory (solid light blue lines). The latter (denoted below by the superscript $L$) is obtained by analogy with Eq.~\eqref{eq:unperturbed_Rg2}, considering that the most relevant long-time effect of the field on the linearized dynamics of the higher-order modes in Eq.~\eqref{eq:eff_lin_dyn_ho_modes} is the correction $\Gamma(0)$ to their relaxation rates $\tilde{\gamma}_j$. This implies that
\begin{equation}
\langle R_g^2 \rangle_{0,\lambda}^{L} =\frac{1}{N} \sum_{n \neq 0} \langle \bm{\chi}^2_n \rangle_{0,\lambda}^{L}=\frac{1}{N} \sum_{n \neq 0} \frac{d\nu T}{ \tilde{\gamma}_n + \Gamma(0) } ,
    \label{eq:Rg_linearized_theory}
\end{equation}
where we introduced the symbol $\langle \dots \rangle_{0,\lambda}^{L}$ to denote the average within the linear-response theory in the absence of any stretching force. We recall here that the linearized theory is only based on the assumption that the Rouse modes are small, and thus it is in principle valid 
%to any order 
at all orders
in the interaction coupling $\lambda$. 
For this reason, upon increasing $\lambda$, it provides an approximation of the actual numerical data which is better than that of the weak-coupling theory, as shown by the two insets of Fig.~\ref{fig:Rg}. 
%%

%%%%%%%%%%%%%%%%%%%%%%%%%%%%%%%%%%%%%
%%%%%%%%%%%%%%%%%%%%%%%%%%%%%%%%%%%%%
\begin{figure}[t]
    \includegraphics[width=0.9\linewidth]{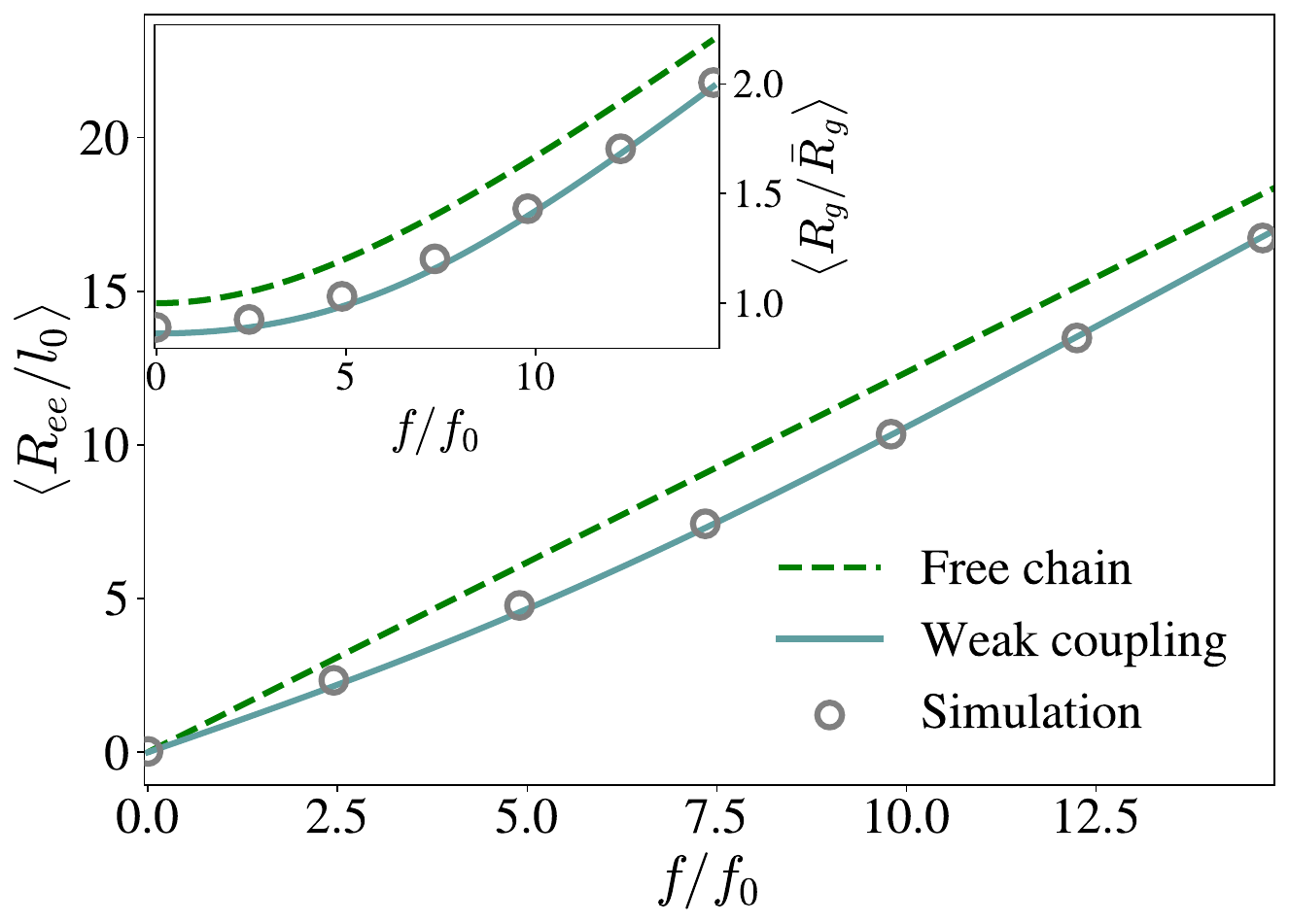}
    \caption{Force-extension curve of a linear polymer with $N=20$ monomers in spatial dimension $d=1$. The main plot shows the response of the average end-to-end distance $R_{\text{ee}}$ to a stretching force $f$. The results of numerical simulations (grey symbols) are compared to the theoretical predictions of Eq.~\eqref{eq:Re_correction} (light blue solid lines). Compared to the free case (i.e., with $\lambda=0$, dashed green line) 
    described by Eq.~\eqref{eq:unperturbed_Ree}, the coupling with the correlated medium introduces an additional resistance to the stretching force due to the attractive field-mediated forces. In this plot, $R_{\text{ee}}$ is measured in units of the typical length $l_0=d T / (\kappa + \kappa_c)$. 
    In the inset we show the behavior of the average gyration radius $R_g$ as a function of the stretching force. Here, $R_g$ is measured in units of $\bar{R}_g$, i.e., of the value it assumes in the free and non-stretched case. 
    %Simulation parameters: $\kappa=1$, $\nu=1$, $T=1$, $D=1$, $R=1$, $\lambda=0.7$. 
    In the simulation we used $\lambda=0.7$, while all other parameters were set to unity.}
    \label{fig:stretching_curve}
\end{figure} 
%%%%%%%%%%%%%%%%%%%%%%%%%%%%%%%%%%%%%
%%%%%%%%%%%%%%%%%%%%%%%%%%%%%%%%%%%%%
%%
%%

\subsection{Force-extension curves}
\label{sec:force-extension}

In this Section we analyze the response of the polymer to the stretching force $\bm{f}_s$, and show how the force-extension curve is modified upon coupling the polymer to the correlated medium. 
The theoretical predictions for the average $\bm R_{\text{ee}}$ and $R_g^2$ reported, respectively, in Eqs.~\eqref{eq:Re_correction} and~\eqref{eq:Rg_correction}, obtained with the weak-coupling approximation, are plotted in Fig.~\ref{fig:stretching_curve} as a function of the modulus $f$ of the stretching force $\bm{f}_s$. In particular, we observe that the correlated fluctuations within the medium effectively increase the stiffness of the polymer, thus introducing an additional resistance to the stretching. Indeed, for all values of $f$, the light blue lines corresponding to the theoretical predictions in Eqs.~\eqref{eq:Re_correction} and \eqref{eq:Rg_correction} and the data obtained via numerical simulations (grey symbols), which are in very good agreement with each other, always lie below the dashed green lines related to the free case (i.e., with $\lambda=0$). This behavior results from the fact that all monomers interact with the field with the same coupling sign $\sigma_i=\sigma$, and thus all field-mediated forces are attractive. 

Although this might not be evident from Fig.~\ref{fig:stretching_curve}, in fact the correction introduced by the coupling to the field vanishes for large stretching forces $f$. This can be evinced by inspecting \cref{eq:Re_correction}, where the integrand function on the r.h.s.~becomes rapidly oscillating for large $f$, so that the integral vanishes.
This can also be rationalized physically by noting that, under an externally imposed stretching,
the typical monomer distance eventually exceeds the range of the field-mediated forces, which thus play a minor role. 
Note that in Fig.~\ref{fig:stretching_curve} the results obtained within the weak-coupling approach are no longer compared with the predictions of the linearized theory (as it was done in Fig.~\ref{fig:Rg}), which relies on the higher-order Rouse modes being 
%sufficiently 
small. 
%\ag{small compared to what?}
%However, 
Indeed,
this assumption is clearly broken when a stretching force is applied, since its effect is that of increasing the typical separation between monomers (and hence the typical size of the Rouse modes via \cref{eq:Rouse_transformation}).

%%%%%%%%%%%%%%%%%%%%%%%%%%%%%%%%%%%%%%%%%%%%%%%%
%%%%%%%%%%%%%%%%%%%%%%%%%%%%%%%%%%%%%%%%%%%%%%%%
\begin{figure*}[t]
    %\centering
    %\begin{subfigure}{0.49\linewidth}
    \subfloat{
        \includegraphics
        %[width=\linewidth, height=6.2cm]
        [width=0.47\linewidth]
        {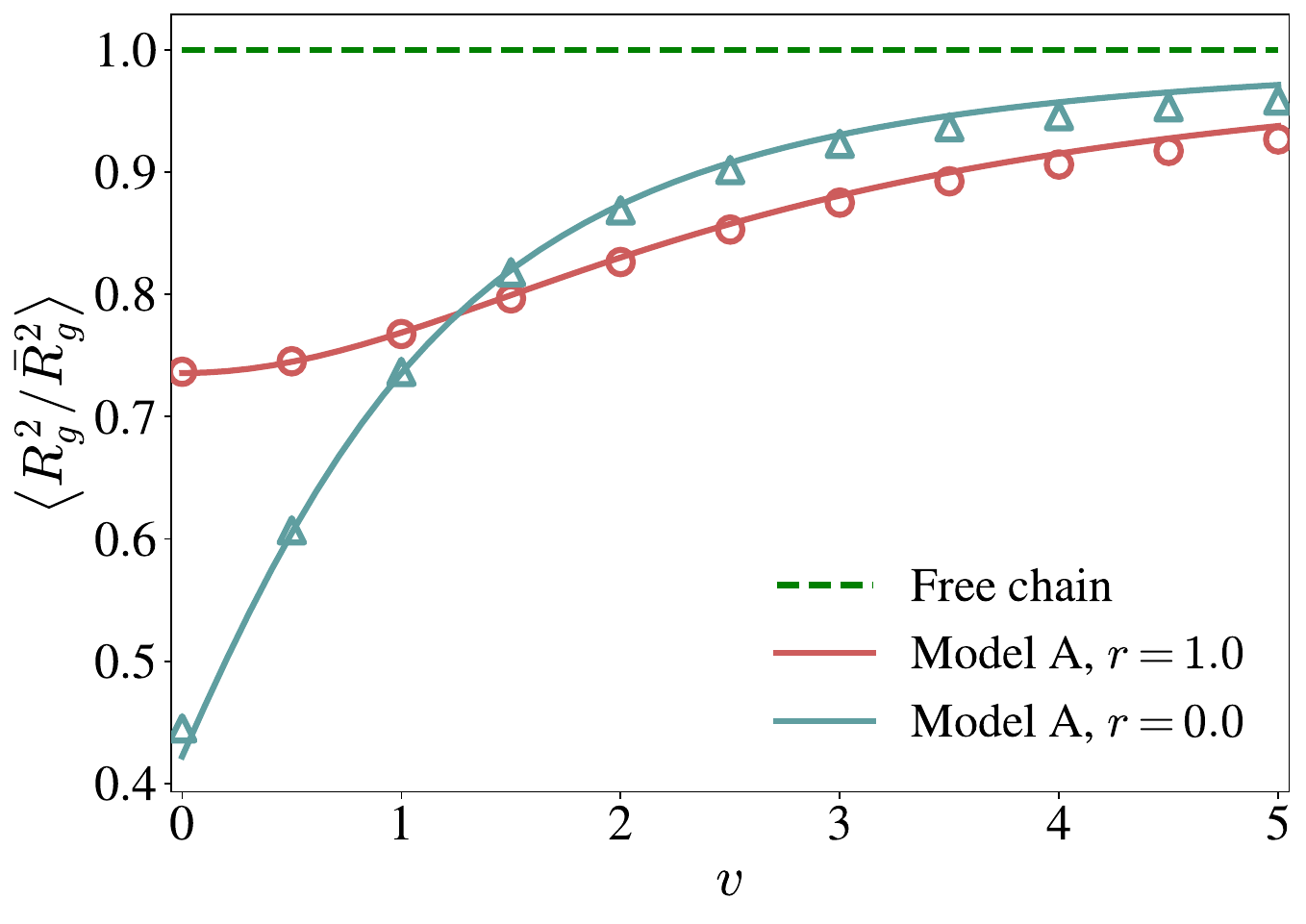}
        \label{moving_trap_a}
    }
    %\end{subfigure}
    %\hfill
    %\begin{subfigure}{0.49\linewidth}
    \subfloat{
        \includegraphics
        %[width=\linewidth, height=6.2cm]
        [width=0.47\linewidth]
        {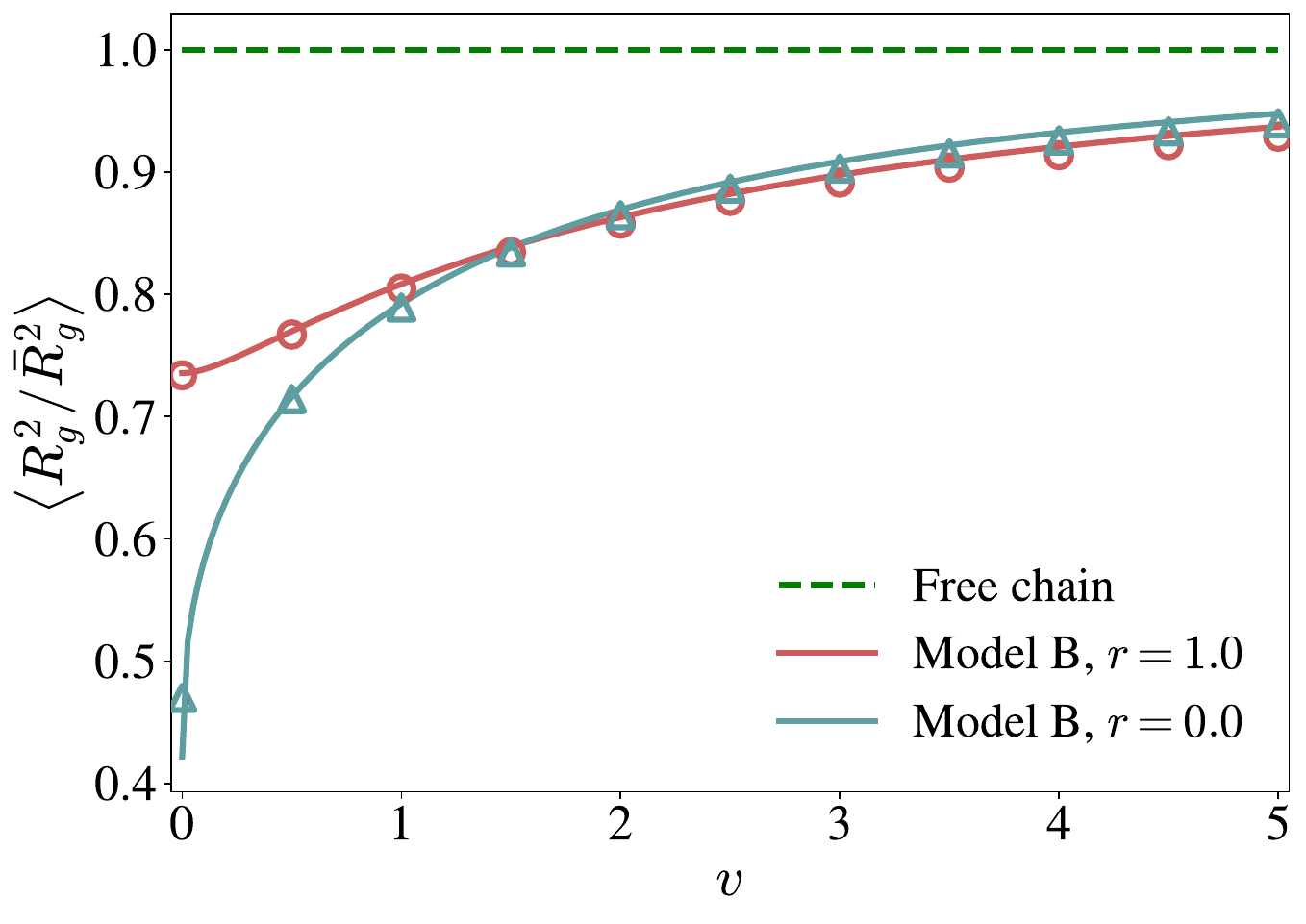}
        \label{moving_trap_b}
    }
    %\end{subfloat}
    \caption{Gyration radius $\langle R_g^2\rangle$ of \rev{a linear} polymer, in the case in which the harmonic confining potential is driven externally at constant velocity $\bm v$ (see \cref{sec:moving-trap}).
    The two panels correspond to model A and model B dynamics of the fluctuating field, respectively. We compare the case without polymer-medium interaction ($\bar R_g^2$, dashed line) to the linear-response prediction for the interacting case (solid lines), 
    for two selected values of the correlation length $\xi_\phi=r^{-1/2}$. Symbols correspond to the results of numerical simulations (see App.~\ref{sec:simulations} for details). In this plot we used $N=10$ and $T=0.1$, while all other parameters were set to unity. 
    }
    \label{fig:moving_trap}
\end{figure*} 
%%%%%%%%%%%%%%%%%%%%%%%%%%%%%%%%%%%%%%%%%%%%%%%%
%%%%%%%%%%%%%%%%%%%%%%%%%%%%%%%%%%%%%%%%%%%%%%%%
%%
%%
%%

\subsection{Typical size out of equilibrium}
\label{sec:moving-trap}

In the previous sections, we have developed a weak-coupling approximation by averaging the observables $R_g^2$ and $\bm R_\T{ee}$ over the equilibrium distribution of the Rouse modes, as in \cref{eq:stat_average}, and expanding it up to its leading order in $\lambda$. This distribution is influenced by the presence of the correlated medium, as encoded in the induced interaction Hamiltonian $\cor H_\T{eff}$ in Eq.~\eqref{eq:eff_Hamiltonian_polymer}. Crucially, however, this distribution does not depend on the \textit{dynamics} of the fluctuating order parameter $\phi(\bm x,t)$ --- as evidenced, for instance, by the fact that $\cor H_\T{eff}$ is by construction the same for both model A and model B dynamics.

This is no longer expected to be the case whenever the system is driven out of equilibrium, for instance via the application of a constant external driving force. To explore this aspect, in this section we assume that the confining harmonic potential to which the monomers are subject is dragged at constant velocity $\bm v$ across the medium, so that the Hamiltonian in \cref{eq:hamiltonian_chain} is modified as 
\begin{equation}
    \mathcal{H}_0=\frac{\kappa}{2} \sum_{ij} M_{ij}\bm{X}_i \cdot \bm{X}_j + \frac{\kappa_c}{2}\sum_{i} (\bm{X}_i-\bm v t)^2.
    \label{eq:hamiltonian_chain-v}
\end{equation}
The corresponding dynamics of the monomers follows from \cref{eq:monomers_dynamics} as 
\begin{align}
    \dot {\bm{X}}_i(t)=&\, -\nu \kappa \sum_j M_{ij}\bm{X}_j - \nu \kappa_c (\bm{X}_i-\bm v t) + \nu \lambda \sigma_i \bm{{\rm f}}(\bm{X}_i) \n\\
    &+ \bm{\xi}_i(t) ,
    \label{eq:monomers_dynamics-v}
\end{align}
which suggests to change coordinates to a comoving reference frame as $\bm Z_i=\bm X_i-\bm v t$; this gives~\cite{venturelli2023memory}
\begin{align}
    \dot {\bm{Z}}_i(t)=&\, -\nu \kappa \sum_j M_{ij}\bm{Z}_j - \nu \kappa_c \bm{Z}_i + \nu \lambda \sigma_i \bm{{\rm f}}'(\bm{Z}_i) + \bm{\xi}_i(t) \n\\
    &-\bm v ,
    %-\nu \kappa \bm v t \sum_j  M_{ij}.
    \label{eq:monomers-Z}
\end{align}
where we used the fact that $\sum_j  M_{ij}=0$ (see \cref{sec:model}).
Above, the form of $\bm{{\rm f}}'$ is the same as that of $\bm{{\rm f}}$ introduced in \cref{eq:field_mediated_forces}, upon replacing the field $\phi$ by its comoving counterpart $\phi'(\bm x,t)= \phi(\bm x+\bm v t,t)$. The response function of the comoving field $\phi'$ is easily found to be~\cite{venturelli2023memory}
\begin{equation}
    G_{\bm{q}}\v (t)=\Theta(t)\mathrm{e}^{-\alpha_{\bm{q}}t+i\bm q\cdot \bm v t}
    \label{eq:propagator_v}
\end{equation}
(compare with \cref{eq:propagator}).
Proceeding as in \cref{sec:model}, we now step to the Rouse modes using (compare with \cref{eq:Rouse_transformation})
\begin{equation}
    \bm{\chi}_i=\sum_{j=0}^{N-1} \varphi_{ij} \bm{Z}_j .
    \label{eq:Rouse_transformation-v}
\end{equation}
In doing so, we focus in particular on the 
%two terms 
term
proportional to $\bm v$ that appears in \cref{eq:monomers-Z}, and note that 
%\plm{leviamo somma in l in (61)?}
\begin{equation}
    \sum_{j=1}^{N-1}\varphi_{ij} 
    %= N^\frac{3}{2}
    \propto
    \delta_{i0},
\end{equation}
where
we used \cref{eq:identity_varphi}.
We thus deduce that the driving $\bm v$ only enters explicitly the equation of motion of $\bm \chi_0$, but not that of the higher-order Rouse modes $\bm \chi_i$ with $i>0$ --- which instead are only affected by $\bm v$ through the response function $G\q\v(t)$ in \cref{eq:propagator_v}.
In particular, this result is independent of the topology of the polymer (encoded in $\varphi_{ij}$).

We are thus in the position to easily compute the gyration radius $\expval{R_g^2}$, which is notably independent of $\bm \chi_0$ (see \cref{eq:def_Rg}). Within the linearized theory, analogous steps lead to the same result as in \cref{eq:Rg_linearized_theory}, upon replacing the memory kernel $\Gamma(t)$ introduced in \cref{eq:lin_mem_kernel_polymer} by 
\begin{equation}
    \Gamma\v(t)=\frac{ND\nu \lambda^2}{d} \int \dslash{q} \,  \frac{q^{2+a} |V_{\bm{q}}|^2 G_{\bm{q}}\v(t)}{\alpha_{\bm{q}}-i\bm q \cdot \bm v} ,
    \label{eq:lin_mem_kernel_polymer_v}
\end{equation}
whence in particular
\begin{equation}
    \Gamma\v(0)=\frac{ND\nu \lambda^2}{d} \int \dslash{q} \,  \frac{q^{2+a} |V_{\bm{q}}|^2 \alpha_{\bm{q}}}{\alpha_{\bm{q}}^2+(\bm q \cdot \bm v)^2}.
\end{equation}
This prediction is tested in \cref{fig:moving_trap}, which shows the average gyration radius as a function of the dragging velocity $\bm v$ (as measured in the stationary state), separately for model A (left panel) and model B (right panel) dynamics of the fluctuating field, and for two selected values of the correlation length $\xi_\phi=r^{-1/2}$. 
For $v=0$ (i.e., at equilibrium), the value of the gyration radius is actually independent of the field dynamics (see \cref{fig:Rg}), and thus it is the same for model A and B. For finite $v$, 
in both cases, 
the correction to the gyration radius due to the field decreases monotonically upon increasing the driving speed $v$. This is consistent with the expectation that, at high $v$, once the field is excited by the interaction with the polymer in a certain region of space, it is no longer able to mediate an interaction between the various monomers, before they are physically transported away by the dragging towards a distinct spatial region.

%%%%%%%%%%%%%%%%%%%%%%%%%%%%%%%%%%%%%%%%%%%%%%

\section{Conclusions}
\label{sec:conclusions}

In 
%the present 
this work we analyzed the behavior of a 
\rev{Rouse}
polymer linearly coupled to a correlated medium described by a fluctuating scalar Gaussian field $\phi(\bm{x},t)$. The reciprocal interaction between the polymer and the field is taken into account in their joint stochastic dynamics, which is assumed to satisfy detailed balance, i.e., to occur at thermal equilibrium. 

Working within the linear-response regime, we 
first studied the relaxation of the center of mass of the polymer towards its rest position in a confining potential. This relaxation turns out to be algebraic at long times, if the fluctuating order parameter $\phi(\bm{x},t)$ supports slow relaxational modes, due either to criticality or to the presence of a local conservation law --- see \cref{sec:relaxation} and \cref{fig:relaxation_com}.
Conversely,
the internal structure of the chain described by higher-order Rouse modes always displays an exponential relaxation, with a typical time scale that is shorter compared to the free case (in the case of monomers which have all the same kind of interaction with the field),
due to the effective attraction induced by the medium (see \cref{fig:relaxation_ho_modes}). 

The spatial range of 
%the
these induced interactions is practically determined by %depends on 
the correlation length of the field. 
Using a weak-coupling approximation, we 
%show 
then showed analytically that the gyration radius of the chain 
decreases
as the field approaches the critical point, at which such a correlation length diverges. 
The theoretical predictions are in very good agreement with numerical simulations (see \cref{fig:Rg}). 
Furthermore, we studied how the typical polymer size depends on the polymerization degree $N$, showing that, after an initial increase for small $N$ due to entropic reasons, the dominant effect of the pairwise-additive, field-induced interactions drives the polymer into a compact globule-like state. Within the weak-coupling approximation, in \cref{sec:force-extension} we then
%we 
analyzed the response of the polymer to a tensile force, observing an enhanced resistance of the polymer 
%to 
against the external stretching, whose origin is again to be attributed to the effective attractions between the monomers induced by the medium (see \cref{fig:stretching_curve}).

Additionally, we analyzed the case in which the system is driven out of equilibrium via the application of an external force, which drags at constant velocity the center of the confining potential to which each monomer is subject. In this case, the field-induced forces on the polymer differ from the static ones, and depend explicitly on the particular dynamics of the order parameter field $\phi$. This carries signatures, for instance, on the resulting gyration radius of the polymer, which we predicted and measured as a function of the driving speed $v$ (see \cref{fig:moving_trap}).

Further extensions of this work may address the steady state of the system in the presence of spatial confinement, where boundary conditions are imposed on the fluctuations of the correlated medium~\cite{Gross_2021,venturelli2022tracer}, which is the typical setting in experimental realizations.
Generalization to higher spatial dimensions, albeit straightforward, would also be relevant in view of applications to real polymeric systems.
Besides, the model presented here
%Even more importantly, the model presented in this work 
opens the possibility of studying more complex dynamical settings,
where the field-induced forces 
%may 
are expected to play a major
%play a 
role in determining the nonequilibrium dynamics of the polymer. For instance, it would be interesting to characterize the dynamical response of the chain to a quenched stretching force, i.e., to a force that is suddenly switched on (or off) at time $t=0$. We expect that slow algebraic relaxations might emerge in 
%those 
the cases where the 
%field 
medium is critical or conserved, and the external force produces an average displacement of the center of mass.
% Moreover, in the case studied here, 

\rev{While in this work we 
focused on the simplest possible polymer model, with the aim of analyzing how its properties are affected by the spatio-temporal correlations of the surrounding medium, 
it would be relevant and
%in the future it would be 
important to extend our study towards more realistic models of polymers, accounting first of all for excluded-volume effects.}

Finally, within the formulation of the model we adopted here, the joint stochastic dynamics of the polymer and the field 
%satisfies 
was chosen to satisfy 
%the detailed balance condition, %and 
detailed balance,
so that (in the absence of external forces) the system is characterized by an equilibrium dynamics. An interesting extension of this work would thus be to consider a polymer chain in an active fluctuating correlated medium, whose fluctuations break detailed balance~\cite{muzzeddu2024migration, ravichandir2025transport}, and characterize its behavior under such non-equilibrium conditions. Similarly, a polymer whose monomers are \textit{active} particles is expected to experience nonequilibrium field-mediated forces when immersed in a fluctuating correlated medium, giving rise to interesting interplays with its typical time and length scales.

\begin{acknowledgments}
We thank Fran\c{c}oise Brochard, Timothy F\"{o}ldes, and Jean-Fran\c{c}ois Joanny for interesting discussions. 
AG acknowledges support from MIUR PRIN project “Coarse-grained description for non-equilibrium systems and transport phenomena (CO-NEST)” n.~201798CZL.
%\dav{Il PRIN di Andrea c'è ancora?}
\end{acknowledgments}

%\onecolumngrid
\appendix

\renewcommand{\thefigure}{A\arabic{figure}} 
\setcounter{figure}{0}

\section{Details of the numerical simulation}
\label{sec:simulations}
All the theoretical predictions derived in the present work are compared with numerical simulations of the stochastic equations of motion~\eqref{eq:monomers_dynamics} and~\eqref{eq:field_dynamics_fourier}. In particular, the stochastic dynamics of the polymer is simulated in real space, whereas the evolution of the fluctuating order parameter $\phi$ is simulated in Fourier space (see, e.g., Refs.~\cite{demery2023non, demery2011perturbative}). 
This requires introducing a momentum cutoff for the modes of the field, given by $q_c=2 \pi n_c/L$ with $L$ the box size, and $n_c$ an integer that determines the number of simulated Fourier modes. Specifically, in $d=1$ the number of modes is $n_c+1$, with wave vectors $q=2 \pi n/L$ and $n \in \{0,1,2,...,n_c\}$. 
Note that the modes $\phi_{\bm{q}}$ with $\bm{q}$ living on the left half line of the momentum space can be automatically obtained as $\phi_{-\bm{q}}=\phi^*_{\bm{q}}$, being $\phi(\bm{x},t)$ a real scalar field. 
In particular, we used $n_c=40$ with box size $L=50$ in all figures but Fig.~\ref{fig:relaxation_com}, where we used instead $n_c=25$ and $L=200$ --- indeed,  in the latter case, capturing the long-time power-law relaxation of the center of mass requires considering larger systems.

The stochastic differential equations~\eqref{eq:monomers_dynamics} and~\eqref{eq:field_dynamics_fourier} are integrated using the standard Euler-Maruyama scheme with integration timestep $\Delta t=0.001$. The discretized dynamics of the polymer is given by: 
\begin{align}
    &\bm{X}_i(t+\Delta t)- \bm{X}_i(t)= \label{eq:discretized_dynamics_polymer}\\&-\Delta t \gamma \sum_j M_{ij}\bm{X}_j(t)-\Delta t \gamma_c \bm{X}_i(t)+ \bm{\xi}_i(t)\n\\&- \Delta t\nu \lambda \sigma_i L^d \sum_{\bm{n}\in S} V_{-\bm{n}}\left( \frac{2 \pi \bm{n}}{L}\right) \phi^R_{\bm{n}}(t) \sin\left(\frac{2 \pi \bm{n}\cdot \bm{X}_i(t)}{L} \right) \n 
    \\&- \Delta t\nu \lambda \sigma_i L^d \sum_{\bm{n} \in S} V_{-\bm{n}}\left( \frac{2 \pi \bm{n}}{L}\right) \phi^I_{\bm{n}}(t) \cos\left(\frac{2 \pi \bm{n}\cdot \bm{X}_i(t)}{L} \right) ,\n 
\end{align}
where the set $S$ is defined as $S = \{-n_c,-n_c+1,...,n_c-1,n_c\}^2$, the noises $\{\bm{\xi}_i(t)\}$ are independent zero-mean Gaussian random variables with standard deviation $\sqrt{2\nu T \Delta t}$, and $\phi_{\bm{n}}^R$ and $\phi_{\bm{n}}^I$ are the real and the imaginary part of the mode $\phi_{\bm{n}}$, respectively. The potential $V_{\bm{n}}$ is given by
\begin{align}
    V_{\bm{n}}&=\frac{(2 \pi R^2)^{-d/2}}{L^d}\int_\mathcal{D} \dd^d \bm{x} \exp \left( -\bm{x}^2/2R^2 - i 2 \pi \bm{n}\cdot \bm{x}/L\right) \n\\&\simeq \frac{1}{L^d} \exp \left[ -\frac{1}{2} \left( \frac{2 \pi}{L}\right)^2 \bm{n}^2 R^2\right],
    \label{eq:potential_discretized}
\end{align}
with integration domain $\mathcal{D}=[-L,L]^d$, and where we assumed that the box size $L$ is much larger than the range of interaction between each monomer and the field, i.e., $L \gg R$. The discretized dynamics of the field is given by
\begin{align}
    &\phi_{\bm{n}}^R(t+\Delta t)-\phi_{\bm{n}}^R(t)=-\Delta t\alpha_{\bm{q}} \phi_{\bm{n}}^R(t)\label{eq:discretized_field_R}\\&
    +\Delta t D \lambda V_{\bm{n}} q^a \sum_j \sigma_j \cos \left( \frac{2 \pi \bm{n}\cdot \bm{X}_j(t)}{L}\right) + \zeta^R_{\bm{n}}(t),\n
\end{align}
and 
\begin{align}
    &\phi_{\bm{n}}^I(t+\Delta t)-\phi_{\bm{n}}^I(t)=-\Delta t\alpha_{\bm{q}} \phi_{\bm{n}}^I(t)\label{eq:discretized_field_I}\\&
    -\Delta t D \lambda V_{\bm{n}} q^a \sum_j \sigma_j \sin \left( \frac{2 \pi \bm{n}\cdot \bm{X}_j(t)}{L}\right) + \zeta^I_{\bm{n}}(t),\n
\end{align}
with $\bm{q}=2 \pi \bm{n}/L$. The noises $\{ \zeta^{R}_{\bm{n}}(t)\}$ and $\{ \zeta^{I}_{\bm{n}}(t)\}$ are zero-mean Gaussian random variables with correlations
\begin{equation}
    \langle \zeta^{R,I}_{\bm{n}}(t) \zeta^{R,I}_{\bm{m}}(s) \rangle=\frac{DT}{L^d} \left( \frac{2 \pi |\bm{n}|}{L}\right)^a \delta(t-s) [\delta_{\bm{n},\bm{m}}\pm \delta_{\bm{n},-\bm{m}}].
    \label{eq:correlation_discretized_noise}
\end{equation}

The results presented in \cref{fig:distance_dimer,fig:Rg,fig:stretching_curve}
correspond to stationary quantities, and are thus obtained by considering a single but long realization (of total duration $\mathcal T=10^3$ in real time units), and by averaging over uncorrelated samples taken along the trajectory.
By contrast, the dynamical relaxation presented in Fig.~\ref{fig:relaxation_com} requires to average over a large number of distinct samples (typically $\simeq  10^8$). To speed up the numerical evaluation, for these latter simulations we used $\Delta t = 0.05$, which we verified not to visibly affect the result.

\begin{figure}
    \centering
    \includegraphics[width=0.9\linewidth]{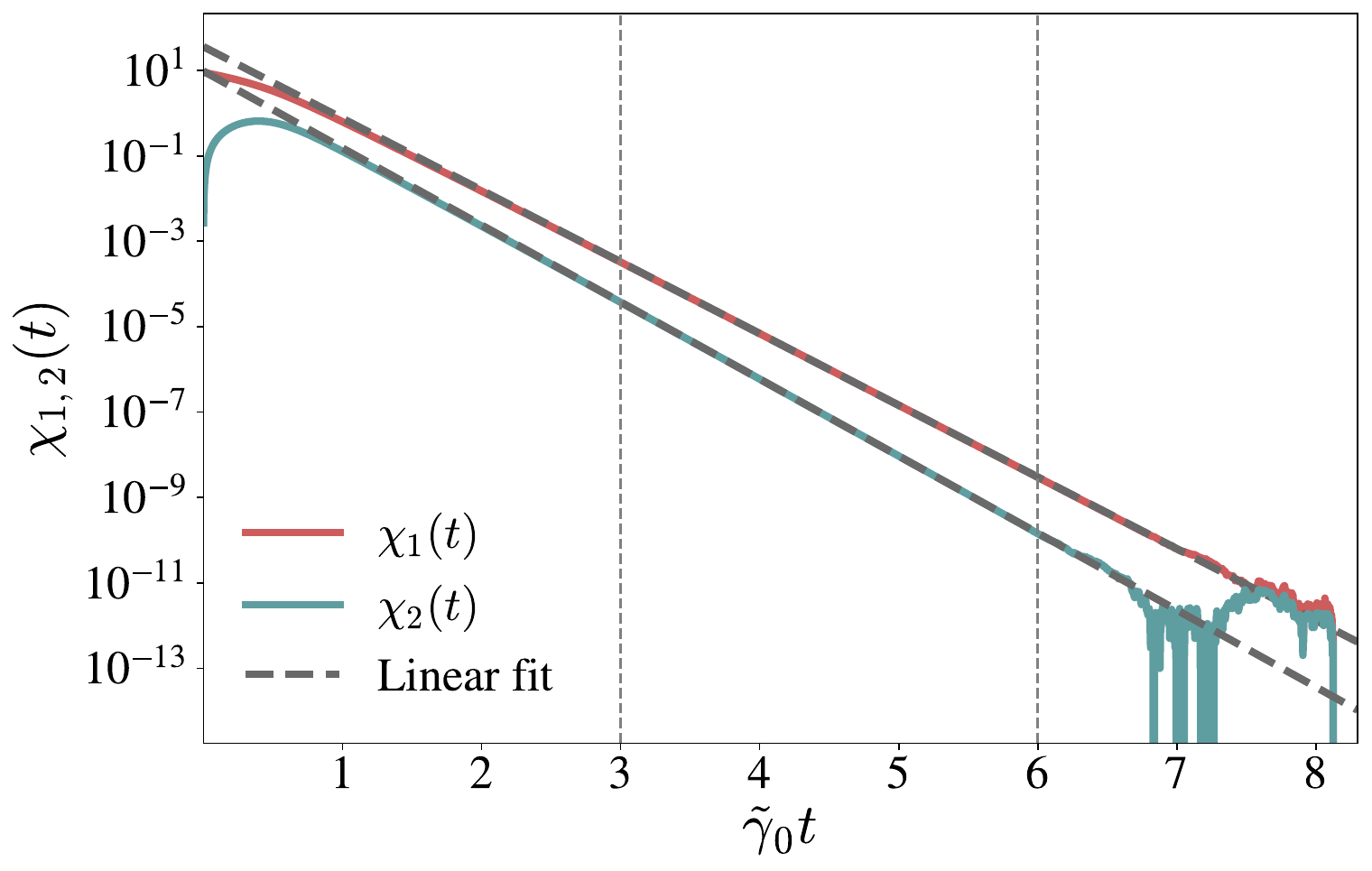}
    \caption{\rev{Example of linear fitting of the 
    temporal decay of
    higher-order Rouse modes, plotted
    %decay 
    on the logarithmic scale. 
    The solid lines correspond to the simulated
    temporal evolution
    %behavior 
    of the first two higher-order Rouse modes. 
    %over time. F
    %(for simplicity, only the first two modes are reported here). 
    Only the 
    numerical data within the
    interval delimited by the vertical dashed lines have been used to perform the linear fitting. In this plot, the correlated field evolved according to critical model B dynamics, while all other parameters are the same as in Fig.~\ref{fig:relaxation_ho_modes}.
    %of the main text.
    } }
    \label{fig:linear_fitting}
\end{figure}

\rev{The relaxation times of the higher-order Rouse modes reported in Fig.~\ref{fig:relaxation_ho_modes} are obtained from a linear fit of the decay of these modes plotted on the logarithmic scale. An example of such a fit is presented in Fig.~\ref{fig:linear_fitting}, in the case where the correlated medium 
%is described by 
evolves according to
critical model B
dynamics. After an initial transient, which depends on the initial condition of both the polymer and the correlated medium, the curves exhibit a linear behavior, corresponding to an exponential decay as a function of time. At long times, however, they become noisy due to the very small amplitude of the modes. For this reason, the linear fit is performed by considering only the numerical data within the time interval delimited by the two dashed vertical lines. }

\section{Long-time behavior of the memory kernel and of the mean position of the center of mass}
\label{app:asymptotic}

In this Appendix we first discuss the asymptotic behavior of the memory kernel $\Gamma(t)$ given in 
%\cref{eq:linear_memory_kernel},
\cref{eq:lin_mem_kernel_polymer},
and secondly its effect on the long-time behavior of the center-of-mass position $\bm \chi_0(t)$ of the polymer (see \cref{sec:relaxation}).

\subsection{Long-time behavior of the memory kernel}
\label{app:asymptotic_kernel}
Let us start from the memory kernel $\Gamma(t)$ in~\cref{eq:lin_mem_kernel_polymer}, 
whose long-time behavior can be easily deduced by inspecting the analytic structure of the corresponding Laplace transform~\cite{venturelli2023memory,hull1955asymptotic}. The latter can be immediately found using \cref{eq:propagator}, and reads
%for which using \cref{eq:propagator} we immediately find
\begin{equation}
    \hat \Gamma(z)= \frac{ND\nu \lambda^2}{d} \int \dslash{q} \,  \frac{q^{2+a} |V_{\bm{q}}|^2 }{\alpha_{\bm{q}}(z+\alpha_{\bm{q}})} .
    \label{appeq:gamma}
\end{equation}
This function is non-analytic for all the points $z \in \mathbb{C}$ that make the denominator vanish, i.e., such that $z+\alpha_q=0$.
Since $\alpha_{\bm{q}}= Dq^a(q^2+r) $, 
%for model B (i.e.~$a=2$) 
at the critical point $r=0$ 
the function $\hat \Gamma(z)$ exhibits a branch cut along the real negative axis in the complex $z$ plane. As there are no other singularities, the behavior of $\hat \Gamma (z)$ around the branching point $z_0=0$ generically determines the long-time behavior of $\Gamma(t)$~\cite{venturelli2023memory}. In particular, if
\begin{equation}
    \hat \Gamma(z) \sim \sum_j a_j (z-z_0)^{\lambda_j}
    \label{appeq:branch_laplace}
\end{equation}
for some $\lambda_j$ around the branching point $z_0$, then
\begin{equation}
    \Gamma(t) \sim \mathrm{e}^{z_0 t} \sum_j \frac{a_j}{\Gamma_E (-\lambda_j) \,t^{1+ \lambda_j}},
    \label{appeq:branch_point}
\end{equation}
as it can be easily understood by taking the inverse Laplace transform term by term. If $\lambda_j$ is a positive integer, then $\Gamma_E (-\lambda_j)$ is not well defined, and the correspondence rather reads 
\begin{equation}
\Gamma(t) \sim a_j \mathrm{e}^{z_0 t}/t^{1+\lambda_j} \leftrightarrow \hat \Gamma (z) \sim a_j \mathrm{e}^{z_0 t} \frac{(-1)^{1+\lambda_j}}{\lambda_j!} s^{\lambda_j} \ln s .
\label{appeq:log}
\end{equation}
If $z_0=0$, \cref{appeq:branch_point} encodes 
%this generates 
an algebraic decay at long times. Note that the very same scenario occurs for the non-critical model B, i.e., $a=2$ and $r>0$, because the branching point is still $z_0=0$.
By contrast, for the non-critical model A (i.e., $a=0$ and $r>0$), the function $\hat \Gamma(z)$ in \cref{appeq:gamma} displays a branch cut on the left of $z_0=-Dr$. Expanding around this new branching point as in \cref{appeq:branch_laplace,appeq:branch_point} will thus generate an exponential prefactor $\mathrm{e}^{z_0 t}=\mathrm{e}^{-Drt}$ superimposed to the algebraic tail.

As we are about to show, expanding $\hat \Gamma(z)$ around the branching point $z_0$ can be easily achieved
even by keeping the interaction potential $V(\vb x)$ generic, under the mild simplifying assumptions that it is rotationally invariant, normalized to unity, and that it depends on a single length scale $R$ (such as in \cref{eq:Gaussian_interaction_potential}). However, for later convenience and to make contact with the notation of Ref.~\cite{venturelli2023memory}, in the following we will not analyze $\hat \Gamma(z)$ directly, but rather introduce $\Gamma(t) = \int_t^{\infty}\dd{u} \mathcal{K}(u)$, so that 
\begin{align}
    &\mathcal{K}(t)= \frac{ND\nu \lambda^2}{d} \int \dslash{q} \,  q^{2+a} |V_{\bm{q}}|^2 G_{\bm{q}}(t) \n\\
    &\mapsto \hat{\mathcal K}(z) =\frac{ND\nu \lambda^2}{d} \int \dslash{q} \,  \frac{q^{2+a} |V_{\bm{q}}|^2 }{z+\alpha_{\bm{q}}},
    \label{appeq:kappa}
\end{align}
and in particular
\begin{equation}
    \hat \Gamma (z) = -[\K (z) -\K (0)]/z.
    \label{appeq:link}
\end{equation}
Note that the function $\K (z)$ in \cref{appeq:kappa} has the same analiticity properties as $\hat \Gamma (z)$, apart from an additional factor $1/z$. 
In the following, we will analyze such properties separately for the cases of critical and non-critical model A and B, which provide access to the corresponding asymptotics of $\Gamma(t)$.

\textit{Critical model A and B. --- }
Let us start from the critical case $r=0$, and expand $\K (z)$ around $z_0=0$.
To get insight on the small-$z$ behavior, it is convenient to
change the integration variable as $y=Dq^{a+2}/z$ in \cref{appeq:kappa} after introducing polar coordinates, which gives~\cite{venturelli2023memory}
\begin{align}
    \K (z)  &= \frac{N \lambda^2 \nu c_d}{2+a} \left(\frac{z}{D}\right)^\frac{d}{2+a} \int_0^\infty \dd{y} \frac{ y^{d/(2+a)}}{1+y} |V_{(zy/D)^\frac{1}{2+a}}|^2 \n\\
    &\sim \frac{N \lambda^2 \nu c_d}{2+a}  \left[ \int_0^\infty \dd{y} \frac{ y^\frac{d}{2+a}}{1+y}\right] \left(\frac{z}{D}\right)^\frac{d}{2+a}.
    \label{eq:kappa_critical}
\end{align}
In the last step we expanded the integrand for small $z$ by using the normalization condition $V_q=1+\order{q^2}$ of the interaction potential, while 
\begin{equation}
    c_d = 2^{1-d}/[d \pi^{d/2} \Gamma_E(d/2)]
\end{equation}
is a numerical constant, accounting for the integration over the angular variables, with $\Gamma_E$ the Euler gamma function.  
Unfortunately, however, the remaining integral over $y$ in \cref{eq:kappa_critical} is divergent at large values of $y$ (i.e., in the UV) --- which is consistent, in hindsight, with the fact that $\hat{\cor K} (0)\neq 0$. This divergence can be cured by improving the expansion in \cref{eq:kappa_critical} as
\begin{align}
    \K (z) \sim \K (0) - \frac{N \lambda^2 \nu c_d}{2+a} \left[ \int_0^\infty \dd{y} \frac{ y^\frac{d}{2+a}}{(1+y)^2}\right] \left(\frac{z}{D}\right)^\frac{d}{2+a}.
    \label{eq:kappa_critical2}
\end{align}
This renders a convergent integral for model B in $d\leq 3$
and for model A in $d<2$, 
whereas for model A in $d>2$ one needs to go one step beyond and write
\begin{align}
    \K (z)\sim &\, \K (0)+z\K'(0) \n\\
    &+ \frac{N \lambda^2 \nu c_d}{2+a} \left[ \int_0^\infty \dd{y} \frac{ y^\frac{d}{2+a}}{(1+y)^3}\right] \left(\frac{z}{D}\right)^\frac{d}{2+a}.
    \label{eq:kappa_critical3}
\end{align}
(For model A in $d=2$, one can check from \cref{appeq:kappa} that $\K'(0)$ diverges logarithmically in the IR, as expected from \cref{appeq:log} --- however, a close inspection confirms that the decay exponent found in, c.f.,~\cref{eq:final-res-critical} can be analytically continued to $d=2$.)
Note that the dependence on $z^{d/(2+a)}$, which essentially follows from dimensional analysis, does not change upon improving this estimate in \cref{eq:kappa_critical,eq:kappa_critical2,eq:kappa_critical3} (whereas the prefactor of the term $z^{d/(2+a)}$ does change).
Using \cref{appeq:link}, we thus deduce that in general, for small $z$,
\begin{equation}
    \hat \Gamma (z) \sim z^{d/(2+a)-1},
\end{equation}
and comparing with \cref{appeq:branch_point,appeq:branch_laplace}
we can conclude that
\begin{equation}
    \Gamma (t) \sim t^{-d/(2+a)}
    \label{eq:final-res-critical}
\end{equation}
at long times.
This corresponds to the result reported in \cref{eq:asymptotics_Gamma_A,eq:asymptotics_Gamma_B}, for $a=0$ and 2, respectively.

\textit{Non-critical model B. --- }
Let us now inspect the non-critical case $r>0$. For model B, i.e., $a=2$, the branching point is still at $z_0=0$, so that using again polar coordinates and changing variables to $y\equiv Dq^2/z$, we find from \cref{appeq:kappa}
\begin{align}
    &\K (z)-\K (0)-z\K'(0)  \n\\
    &\sim \frac{N\lambda^2 \nu c_d}{2} \int_0^\infty \dd{y} \frac{ y^{1+\frac{d}{2}}}{(1+yr)^3}  \left(\frac{z}{D}\right)^{1+\frac{d}{2}},
    \label{appeq:kappa_off_modelB}
\end{align}
for $d<2$, and 
\begin{align}
    &\K (z)-\K (0)-z\K'(0) -\frac{z^2}{2} \K''(0) \n\\
    &\sim -\frac{N\lambda^2 \nu c_d}{2} \int_0^\infty \dd{y} \frac{ y^{1+\frac{d}{2}}}{(1+yr)^4}  \left(\frac{z}{D}\right)^{1+\frac{d}{2}},
    \label{appeq:kappa_off_modelB2}
\end{align}
for $d>2$.
Using \cref{appeq:branch_point,appeq:branch_laplace,appeq:link} we thus get
\begin{equation}
    \hat \Gamma (z)\sim z^{d/2} \quad \longrightarrow \quad \Gamma(t) \sim t^{-(1+d/2)},
    \label{appeq:res-off-B}
\end{equation}
as reported in \cref{eq:asymptotics_Gamma_A}. (Again, evaluating the prefactor for the case $d=2$ requires to inspect the logarithmic divergence of $\K'(z)$ for small $z$, whereas the estimate of the algebraic decay exponent in \cref{appeq:res-off-B} turns out to apply also to $d=2$.)

\textit{Non-critical model A. --- }Finally we address the non-critical case of model A. The corresponding branching point is at $z_0=-Dr$, and expanding $\K (z)$ around it yields
\begin{align}
    \K (z) &= N\lambda^2 \nu D c_d \left(\frac{z+Dr}{D} \right)^\frac{d}{2}\int_0^\infty \!\! \dd{y} \frac{ y^{d/2}}{1+y}  |V_{\sqrt{y(z/D+r)} }|^2\n\\
    &\sim (z+Dr)^\frac{d}{2},
   \label{appeq:kappa_off_modelA}
\end{align}
where we called $y=Dq^2/(z+Dr)$, and in the last step we expanded in small powers of $(z+Dr)$. By contrast, note that by expanding the same expression around $z=0$, one  would get 
%$\K (z)= \K (0)  +\order{z^{d/2}}$, meaning that for any $d\geq 1$ 
$\K (z)= \K (0)  +\order{z}$, meaning that
the function $\hat \Gamma(z)$ in \cref{appeq:link} is \textit{not} singular in $z=0$. The singularity in $z=z_0=-Dr$ thus still dominates the long-time asymptotics of $\Gamma(t)$, which follows from \cref{appeq:branch_point,appeq:branch_laplace,appeq:link} as
\begin{equation}
    \hat \Gamma (z)\sim (z+Dr)^{d/2} \quad \longrightarrow \quad \Gamma(t) \sim \mathrm{e}^{-Dr t} t^{-(1+d/2)},
\end{equation}
as reported in \cref{eq:asymptotics_Gamma_A}.

\subsection{Dynamics of the mean center-of-mass position}
\label{app:asymptotic_com}

The evolution of the mean position $\expval{\chi_0^\alpha (t)}$ of the center of mass of the polymer, starting from the initial condition $\bar{\chi}_0$
at time $t_0=0$, is given in the Laplace domain by \cref{eq:GLE_com_dimer_laplace}.
The latter can be rewritten, in terms of the function $\K (z)$ introduced in \cref{appeq:kappa}, as 
\begin{equation}
\hat{\chi}^\alpha_0(z)=\frac{\bar{\chi}_0}{z+\tilde{\gamma}_0-[\K(z)-\K(0)]}.
    \label{eq:GLE_com_dimer_NJP}
\end{equation}
This function exhibits two types of singularities in the complex $z$ plane: (i)~a branch cut starting from the branching point $z_0$, where $\K(z)$ is non-analytic, as discussed in the previous section; and 
%(ii)~a pole in $z=z^*$, 
(ii)~the zero(s) $z=z^*$ of the denominator $\mathcal D(z)$, 
implicitly defined by the condition
\begin{equation}
    \mathcal D(z^*) =     z^*+\tilde{\gamma}_0-[\K(z^*)-\K(0)]  \equiv 0.
    %z^*= -\tilde{\gamma}_0+[\K(z^*)-\K(0)].
    \label{appeq:pole_dressing}
\end{equation}
Note that these are simple poles: to see this, it is sufficient to take the derivative
\begin{equation}
    \mathcal D'(z) = 1-\K'(z) = 1+\frac{ND\nu \lambda^2}{d} \int \dslash{q} \,  \frac{q^{2+a} |V_{\bm{q}}|^2 }{(z+\alpha_{\bm{q}})^2},
\end{equation}
where we used \cref{appeq:kappa}, and note that $\mathcal D'(z^*)$ cannot vanish.
Moreover, note that it must be that $\Re z^*<0$ and $\Re z_0<0$ in order for $\expval{\chi_0^\alpha (t)}$ to decay to zero at long times.
The closest to the imaginary axis $\Re z=0$ among $z_0$ and $z^*$ generically determines the long-time asymptotic behavior of $\expval{\chi_0^\alpha (t)}$ \cite{venturelli2023memory,hull1955asymptotic}. 

Again, it proves convenient to inspect
%the cases of model A/B separately.
first the critical case $r=0$, for which the branching point $z_0=0$ necessarily dominates the
long-time asymptotics of $\hat{\chi}^\alpha_0(z)$
in \cref{eq:GLE_com_dimer_NJP}.
%is then found by e
Expanding the latter around
$z_0=0$ gives
\begin{align}
    \hat{\chi}^\alpha_0(z) = \bar{\chi}_0 \sum_{n=0}^\infty  
    %\left[ \K (z) -z  \right]^n    \left[ \K (0) +\tilde \gamma_0  \right]^{-(n+1)}
    \left[ \K (z)-\K (0)  -z  \right]^n    \tilde \gamma_0^{-(n+1)} 
    \sim  z^{d/(2+a)},
    \label{eq:X(s)_geometric}
\end{align}
where in the last step we used \cref{eq:kappa_critical2,eq:kappa_critical3}. Comparing with \cref{appeq:branch_point,appeq:branch_laplace} then yields
\begin{equation}
    \expval{\chi^\alpha_0 (t)} \sim t^{-1-d/(2+a)},
\end{equation}
as reported in \cref{eq:relax_modelA,eq:relax_modelB}.

The situation is analogous for the non-critical model B, i.e., $a=2$ and $r>0$, as the branching point is still $z_0=0$. Using the asymptotic behavior of $\K (z)$ given in \cref{appeq:kappa_off_modelB,appeq:kappa_off_modelB2}
in the second step of \cref{eq:X(s)_geometric} this time gives
\begin{equation}
    \hat{\chi}^\alpha_0(z) \sim z^{1+d/2} \quad \longrightarrow \quad \expval{\chi^\alpha_0 (t)} \sim t^{-2-d/2},
\end{equation}
as stated in \cref{eq:relax_modelB}.

The non-critical case in model A (i.e.~$r>0$ and $a=0$) is more delicate, because the relative positions of the branching point $z_0=-Dr$ and the 
%singular point(s)
pole(s) 
$z^*$ may change depending on the value of $r$.
For sufficiently small $r\ll \tilde\gamma_0/D$, the branching point $z_0$ is closer to the imaginary axis than the 
%singular point(s) 
pole(s) $z^*$. The leading asymptotic behavior of $\chi^\alpha_0(t)$ can thus be found by expanding $\hat{\chi}^\alpha_0(z)$ in \cref{eq:GLE_com_dimer_NJP} around $z_0=-Dr$. Using \cref{appeq:kappa_off_modelA} gives
\begin{equation}
    \hat{\chi}^\alpha_0(z) \sim (z+Dr)^{d/2},
\end{equation}
and comparing with \cref{appeq:branch_point,appeq:branch_laplace} then yields
\begin{equation}
    \expval{\chi^\alpha_0 (t)} \sim \mathrm{e}^{-Dr t} t^{-(1+d/2)},
\end{equation}
as reported in \cref{eq:relax_modelA}.
%%
%%
%%
%%%%%%%%%%%%%%%%%%%%%%%%%%%%%%%%%%
%%%%%%%%%%%%%%%%%%%%%%%%%%%%%%%%%%
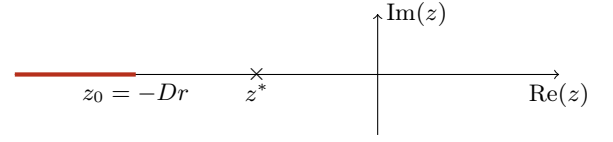
\begin{figure}
\centering
\begin{tikzpicture}[scale=0.8] % Adjust scale to fit in a single column
    % Draw the axes
    \draw[->] (-6, 0) -- (3, 0) node[anchor=north] {Re$(z)$};
    \draw[->] (0, -1) -- (0, 1) node[anchor=west] {Im$(z)$}; % Make the vertical axis shorter

    % Define positions for z^* and -Dr
    \coordinate (zstar) at (-2, 0);
    \coordinate (Dr) at (-4, 0);

    % Draw the pole at z^*
    \node at (zstar) {$\times$};
    \node[below] at (zstar) {$z^*$};

    % Draw the branch cut starting at -Dr and going left
    \draw[ultra thick, BrickRed] (Dr) -- (-6, 0);
    \node[below] at (Dr) {$z_0=-D r$};
\end{tikzpicture}
\caption{Sketch of the analytic structure in the complex $z$ plane of the function $\hat{\chi}^\alpha_0(z)$ in \cref{eq:GLE_com_dimer_NJP}, for the non-critical case of model A, and for $r\gg \tilde\gamma_0/D$. 
This features a branch cut (red) terminating in $z=z_0$, and a simple real pole in $z=z^*$, which is a root of \cref{appeq:pole_dressing}. The latter dominates the long-time asymptotics of $\expval{\chi^\alpha_0(t)}$,
as described in \cref{app:asymptotic_com}.}
\label{fig:poles_modelA}
\end{figure}
%%%%%%%%%%%%%%%%%%%%%%%%%%%%%%%%%%
%%%%%%%%%%%%%%%%%%%%%%%%%%%%%%%%%%
%%
%%
%%
In the opposite limit $r\gg \tilde\gamma_0/D$,
the analytic structure of the function $\hat{\chi}^\alpha_0(z)$ in \cref{eq:GLE_com_dimer_NJP} 
can be inspected by plotting its real or imaginary part using e.g., \texttt{Wolfram Mathematica}, for selected choices of the interaction potential $V_q$. By choosing a Gaussian $V_q$ as in \cref{eq:Gaussian_interaction_potential} and for $d=1$, 2, and 3 (but also for a Yukawa-like potential $V_q=1/(1+q^2R^2)$~\cite{venturelli2023memory}), the resulting analytic structure resembles the one sketched in \cref{fig:poles_modelA}, with a single real pole $z^*$ to the right of the branch cut. 
This simple pole dominates the long-time behavior of $\expval{\chi^\alpha_0 (t)}$, which thus generically reads~\cite{venturelli2023memory,hull1955asymptotic}
\begin{equation}
    \expval{\chi^\alpha_0 (t)} \sim \mathrm{e}^{z^* t},
    \label{appeq:relax_exp}
\end{equation}
with $z^*<0$ given implicitly from \cref{appeq:pole_dressing}.

We remark that the asymptotic behavior of the center of mass $\chi^\alpha_0 (t)$ studied here coincides with that of a single particle at position $X(t)$ relaxing towards the center of a harmonic potential of stiffness $\tilde \gamma_0$, which was derived in Ref.~\cite{venturelli2022nonequilibrium} within a weak-coupling approximation for small $\lambda$, 
%(see Eqs.~(33) and (34) therein, and the Erratum), % only after the Erratum has been sent
and later in Ref.~\cite{venturelli2023memory} within linear response theory (see Appendix~C therein). In particular, for the non-critical model A it was found perturbatively in Ref.~\cite{venturelli2022nonequilibrium} that $X(t)\sim \lambda^2  t \mathrm{e}^{-\tilde \gamma_0 t}+\order{\lambda^4}$, for $Dr>\tilde \gamma_0$~\footnote{\rev{See Eq.~(30) in Ref.~\cite{venturelli2022nonequilibrium}, 
where a prefactor $t$ was inadvertently missing from the first line.}}.
%Unfortunately, a prefactor $t$ was missing from the first line of Eq.~(30) in Ref.~\cite{venturelli2022nonequilibrium}.
In hindsight, this is compatible with the behavior found above in \cref{appeq:relax_exp}, as it can be checked
by using that for small $\lambda$ (see \cref{appeq:pole_dressing})
\begin{equation}
    z^* = -\tilde \gamma_0 +\lambda^2 \left[\K(-\tilde \gamma_0)-\K(0)  \right] +\order{\lambda^4}.
\end{equation}

Finally, we note that the techniques presented in this Appendix can in principle be used to predict the prefactor of $\expval{\chi^\alpha_0 (t)}$ in the long-time limit. Estimates for this prefactor have however not been reported in the main text, because they turn out to potentially depend on the choice of the initial condition of the field $\phi_{\bm{q}}(t_0)$ (encoded in the term $\bar{\bm{v}}_j^{\text{ic}}(t)$ in \cref{eq:linearized_ic_polymer}, and which we fixed in \cref{eq:initial_condition_field} to simplify the calculation). Although a perturbative calculation proves $\phi_{\bm{q}}(t_0)$ to be irrelevant for the long-time dynamics of $\expval{\chi^\alpha_0 (t)}$ at leading order for small coupling $\lambda$, numerical evidence shows, instead, that the prefactor of the algebraic decay can get \textit{dressed} when $\lambda$ becomes larger. These corrections, whose magnitude depends on $\phi_{\bm{q}}(t_0)$, do not alter the algebraic decay exponents of $\expval{\chi^\alpha_0 (t)}$, which are correctly captured both by the perturbative calculation in Ref.~\cite{venturelli2022nonequilibrium}, and by the linear-response calculation presented here (although for a selected choice of $\phi_{\bm{q}}(t_0)$).

\section{Perturbative correction to the gyration radius}
\label{app:corr_Rg}

In this Appendix we derive the expression of the generating functional $\mathcal{Z}[\{ \vb j_i\}]$ reported in Eq.~\eqref{eq:generating_functional}, and we use it to predict analytically the typical polymer size in Eqs.~\eqref{eq:Rg_correction} and~\eqref{eq:Re_correction} within the framework of the weak-coupling approximation. By taking appropriate derivatives of $\mathcal{Z}[\{ \vb j_i\}]$, one can compute the covariance between the $R_g^2$ (or $\bm{R}_{\text{ee}}$) and the effective Hamiltonian $\mathcal{H}_{\text{eff}}$, which constitute the first non-trivial corrections of order $\lambda^2$ to the unperturbed values $\langle R_g^2\rangle_{f,0}$ and $\langle \bm{R}_{\text{ee}}\rangle_{f,0}$. After rewriting the Hamiltonian $\mathcal{H}_0$ 
in \cref{eq:hamiltonian_chain}
in terms of the Rouse modes $\{ \bm{\chi}_i\}$ as
\begin{equation}
    \mathcal{H}_0=
    %\frac{\kappa}{2} \sum_{ij} M_{i j} \bm{X}_i \cdot \bm{X}_j + \frac{\kappa_c}{2} \sum_{i} \bm{X}_i^2=
    \frac{1}{2} \sum_i \mathcal{M}_i \bm\chi_i^2 ,
    \label{eq:H0_rouse_modes}
\end{equation}
we use the definition of the generating functional given in Eq.~\eqref{eq:generating_functional} to obtain:
\begin{align}
    &\mathcal{Z}[\{ \vb j_i\}]=\Big \langle \exp \Big(\sum_{i} \vb j_i \cdot \bm{\chi}_i  \Big) \Big \rangle_{f,0} \\&=\frac{1}{\mathcal{N}} \prod_{j=0}^{N-1} \int \dd \bm{\chi}_j \mathrm{e}^{ -\frac{\beta}{2} \mathcal{M}_i \bm\chi_j^2 - \bm{\chi}_j \cdot \left[\beta (\varphi_{j,N-1}-\varphi_{j,0})-\vb j_j \right]}\n\\&
    =\exp \left[ \frac{1}{2\beta} \sum_i \frac{1}{\mathcal{M}_i} [\vb j_i^2 + 2 \beta (\varphi_{i,N-1}-\varphi_{i,0})\bm{f}_s\cdot \vb j_i]\right], \n
    \label{eq:der_gen_fun}
\end{align}
where $\mathcal{N}$ denotes the normalization factor given by $\mathcal{N}=\mathcal{Z}[\{ \vb j_i=\bm{0}\}]$, and the integral in the second line can be easily solved, being a standard multivariate Gaussian integral. In order to compute the covariance between $R_g^2$ and $\mathcal{H}_{\text{eff}}$, we first 
%of all 
need to evaluate $\langle \mathcal{H}_{\text{eff}} \rangle_{f,0}$, which depends on the following averages:
\begin{equation}
    \Big \langle \exp \Big(i \bm{q} \sum_{k} (\varphi_{ki}-\varphi_{kj})\bm{\chi}_k \Big) \Big \rangle=\mathcal{Z}[\{ \vb j_k=i \bm{q} (\varphi_{ki}-\varphi_{kj})\}] , \label{eq:der_gen_fun_2}
\end{equation}
with generic indices $i$ and $j$. Thus we have:
\begin{align}
    &\langle \mathcal{H}_{\text{eff}} \rangle_{f,0}=
    \label{eq:der_gen_fun_3}\\&-\frac{\lambda^2}{2} \sum_{i j}\int  \dslash{q}  \abs{V_q}^2 \cor C\q \left[\mathcal{Z}[\{ \vb j_k=i \bm{q} (\varphi_{ki}-\varphi_{kj})\}]-1\right]. \n
\end{align}
Secondly, we need the correlation between the gyration radius and the effective Hamiltonian, i.e.~$\langle R_g^2 \mathcal{H}_{\text{eff}} \rangle_{f,0}$. Since the $R_g^2$ 
%is
features a weighted combination of the Rouse amplitudes $\bm{\chi}_j^2$, such correlation requires the knowledge of the following average:
\begin{align}
    &\langle \bm \chi_n^2 \mathrm{e}^{i\vb q \cdot \sum_k ( \varphi_{ki}-\varphi_{kj})\bm \chi_k}\rangle_{f,0} = \sum_{\alpha} \eval{\frac{\partial^2 \mathcal Z[\{ \vb j_k\}]}{\partial j_n^\alpha \partial j_n^\alpha}}_{\vb j_k = i \vb q (\varphi_{ki}-\varphi_{kj})} \n\\
    &=\left\lbrace\frac{d}{\beta \mathcal{M}_n} + \left[ \frac{i\bm{q}/\beta (\varphi_{ni}-\varphi_{nj})+\bm{f}_s(\varphi_{n,N-1}-\varphi_{n,0})}{\mathcal{M}_n}\right]^2\right\rbrace \n\\
    &\quad \,\times \mathcal{Z}[\{ \vb j_k=i \bm{q} (\varphi_{ki}-\varphi_{kj})\}].
    \label{eq:der_gen_fun_4}
\end{align}
Combining Eqs.~\eqref{eq:pert_exp}, \eqref{eq:unperturbed_Rg2}, \eqref{eq:der_gen_fun_3} and \eqref{eq:der_gen_fun_4}, we get the first non-trivial perturbative correction to the gyration radius $R_g^2$ reported in Eq.~\eqref{eq:Rg_correction} of the main text. In the case of the $\bm{R}_{\text{ee}}$, being the latter a linear combination of the Rouse modes as shown in~\eqref{eq:def_Ree}, its correlation $\langle \bm{R}_{\text{ee}} \mathcal{H}_{\text{eff}} \rangle_{f,0}$ with the effective Hamiltonian depends on the average:
\begin{align}
    &\langle \chi^\alpha_n \mathrm{e}^{i\vb q \cdot \sum_k ( \varphi_{ki}-\varphi_{kj})\bm \chi_k} \rangle_{f,0}=\eval{\frac{\partial \mathcal Z[\{ \vb j_k\}]}{\partial j_n^\alpha }}_{\vb j_k = i\bm{q}(\varphi_{ki}-\varphi_{kj})} 
     \n \\&=\frac{{\rm j}_n^\alpha + \beta f_s^\alpha (\varphi_{i,N-1}-\varphi_{i,0})}{\beta \mathcal{M}_n} \mathcal{Z}[\{ \vb j_k=i \bm{q} (\varphi_{ki}-\varphi_{kj})\}].
\end{align}
The equation above can be combined with Eqs.~\eqref{eq:pert_exp},~\eqref{eq:unperturbed_Ree}, and~\eqref{eq:der_gen_fun_3} to obtain the result reported in Eq.~\eqref{eq:Re_correction} of the main text.

\bibliography{references}

\end{document}